\documentclass{aa}
\usepackage[utf8]{inputenc}
\usepackage{natbib}
\usepackage[english]{babel}
\usepackage{graphicx}
\usepackage{amsmath}
\usepackage{MR_AA}

\begin{document}

\title{Dynamics of the envelope of a rapidly rotating star or giant
planet in gravitational contraction}

\author{D. Hypolite\inst{1,2}
\and M. Rieutord\inst{1,2}}

\institute{Universit\'e de Toulouse; UPS-OMP; IRAP; Toulouse, France
\and CNRS; IRAP; 14, avenue Edouard Belin, F-31400 Toulouse, France}

\date{\today}

\abstract{
}{We wish to understand the processes that control the fluid flows of a
gravitationnally contracting and rotating star or giant planet.}
{We consider a spherical shell containing an incompressible fluid that
is slowly absorbed by the core so as to mimick gravitational
contraction. We also consider the effects of a stable stratification
that may also modify the dynamics of a pre-main sequence star of
intermediate mass.}
{This simple model reveals the importance of both the Stewartson
layer attached to the core and the boundary conditions met by the fluid
at the surface of the object. In the case of a pre-main sequence star of
intermediate mass where the envelope is stably stratified, shortly after
the birth line, the spin-up flow driven by contraction overwhelms the
baroclinic flow that would take place otherwise.  This model also shows
that for a contracting envelope, a self-similar flow of growing amplitude
controls the dynamics. It suggests that initial conditions on the birth
line are most probably forgotten. Finally, the model shows that the shear
(Stewartson) layer that lies on the tangent cylinder of the core is likely
a key feature of the dynamics that is missing in 1D models. This layer
can explain the core and envelope rotational coupling that is required
to explain the slow rotation of cores in giant and subgiant stars.
}{
}


\authorrunning{Hypolite \& Rieutord}
\titlerunning{Dynamics of the envelope of a rotating and contracting...}

\maketitle

\section{Introduction}

The influence of rotation has long been known to be crucial for understanding
mixing in radiative regions of stars and interpreting the observed surface
abundances \cite[][]{strittmatter69}. Many (if not all) stellar evolution
codes now include some modelling of the transport of angular momentum
and chemical elements by the flows induced by rotation in the stably
stratified radiative zones. The difficulty is that stellar
evolution code are one-dimensional while fluid flows are generally
multi-dimensional. Basically, two types of modelling are currently
used: one is based on a simple (turbulent) diffusion process
\cite[e.g.][]{pinsonneault97} and the other includes a first-order
modelling of meridional advection, distinguishing the transport of
angular momentum and that of chemicals \cite[][]{zahn92}. However,
both of them need adjustment of diffusion coefficients with respect
to observations. While this modelling has succeeded in explaining
various features of abundances pattern or evolutionary effects like the
surface abundance of lithium as a function of mass \cite[][]{CT99},
the (relative) high number of red super-giants in low-metallicity
galaxies \cite[][]{MM01}, or the ratio of type Ibc to type II supernovae
\cite[][]{MM05}, many recent results challenge our understanding of this
so-called rotational mixing. One of the most famous is the distribution
of LMC B-stars in a diagram plotting the nitrogen abundance versus
the rotational velocity (the so-called Hunter diagram). As shown
by \cite{brott_etal11}, many slowly rotating stars show overabondance
of nitrogen compared to the predictions of models while some fast rotating
stars show much less nitrogen than expected.

In order to progress  in this difficult problem, it is clear that a
better view of the dynamics of rotating stars is needed. Many processes
contribute to rotational mixing. Let us recall that the time scale of
element transport inside stars is essentially controlled by the radiative
zones. Indeed, convective regions mix almost instantaneously. When the
star is non-rotating elements may migrate through radiative regions
thanks to microscopic diffusive processes or propagating waves. In
rotating stars radiative regions are no longer at rest: beyond the
global rotation baroclinic flows arise as a differential rotation and
an associated meridional circulation \cite[][]{R06b}. While meridional
currents can obviously carry elements from deep to surface layers,
differential rotation can also contribute to the transport through
shear driven instabilities and associated turbulence. As pointed out by
\cite{zahn92}, baroclinicity may be helped by angular momentum losses
resulting from mass loss. To be complete, a sequence of gravitational
contraction may also drive a redistribution of angular momentum and
elements during the pre-main sequence or just at the end of the main
sequence.

Going beyond the above mentioned modelling of rotational mixing requires
relaxing the spherical symmetry of the models. A first step in this
direction is to simplifies the structure of the stars so as to focus
on its interior hydrodynamics. This line of research was followed in
\cite{R06}. The star was reduced to a spherical ball filled with an
incompressible fluid in order to study the properties of the baroclinic
flows. Besides giving a simplified view of the dynamic processes
controlling a radiative zone, this work lead to a simplified set-up for
boundary conditions in complete (compressible) two-dimensional models of
fast rotating stars \cite[][]{ELR13}.  In the present paper we address the
effect of mass-contraction/expansion of a star or a giant planet on the
global flows that affect the envelope of the rotating object. Although
this driving results from the compressibility of the fluid, we decided
to investigate its consequences using an incompressible fluid. The
contraction/expansion of the envelope is mimicked by a core that
absorbs or injects matter at a constant rate in a surrounding stably
or neutrally stratified envelope. We wish to determine the resulting
flows and the circumstances when it overwhelms the eventual baroclinic
flows. This will complete the work of \cite{RB14} who studied this same
competition in the case of a spin-down driven by mass-losses.

The paper is organized as follows : In section 2 we describe the model
{and the questions of fluid dynamics that are addressed}.  We next consider
the case where outer boundary conditions are no-slip and use the ensuing
analytical solutions to enlight the dynamics (section 3). In section 4,
we discuss the presumably more realistic case with stress-free boundary
conditions. Conclusions and discussion of extrapolations of the results
to models with a compressible fluid follow.

\section{The model}

\subsection{Description}

For modelling in a simple way the contracting/expanding star or giant planet, we
consider a self-gravitating incompressible viscous fluid enclosed between
two spherical shells.  These shells are assumed not to be distorted
by rotation.  The inner shell may represents a core-envelope boundary
through which a flux of matter is imposed. This flux is described by
a uniform radial velocity $V_{s}$ at this interface. For a contracting
envelope, $V_s$ is negative. Note that as the fluid of the envelope
enters the core and as its radius is assumed fixed, the core's density
linearly increases with time, namely

\[ \rho_{\rm core} = \rho_0\lp1+\frac{t}{t_s}\rp\]
where we introduced the initial density of the core $\rho_0$ and the
``{suction} time", the time the core needs to increase its density by a
factor $2$, namely

\begin{equation}
t_{s}=\frac{R_{\rm core}}{3V_{s}} \tilde\rho
\label{tmsmodel}
\end{equation}
where $\tilde\rho=\frac{\rho_0}{\rho}$. This time is of same order of
magnitude as the Kelvin-Helmholtz time. Here, $\rho$ is the density of
the envelope. We note that changing the sign of $V_{s}$ can readily describe
an expanding envelope due to a wind. The outer bounding sphere is of
constant radius and lets matter going through so as to insure mass
conservation in the envelope.

\subsection{A digest of the following Fluid Dynamics}

The following of the paper is basically fluid dynamics. We thought that
the reader interested in the astrophysical implications of the results
might also be interested in a summary of the various steps that are
described below. The reader may then jump to the discussion
at the end of the paper.

As it is well known, the contraction of the star together with the
conservation of angular momentum induces a global acceleration of the
rotation rate, namely a spin-up, of the star.
The problem of spin-up flows has been considered many times in the fluid
dynamics literature \cite[see the review of][]{duck_etal01}. However,
these studies have been mostly motivated by engineering applications and
therefore have considered a driving by boundaries and not, as in our case, by a
radial flow. The influence of a stable stratification, that we need to
know in order to deal with radiative regions of stars, has been more
seldom considered in astrophysically relevant geometries. The most
relevant study is certainly the work of \cite{fried76} who considered the
spin-down of the radiative core of the Sun through Reynolds stresses at
the interface of the convective and radiative regions. So, here too, the
driving of the flows is by the boundaries. Friedlander's model is
much simplified (as ours) as it uses the Boussinesq approximation
(i.e. neglecting the compressibility of the fluid). It also neglects
baroclinic flows associated with this set-up. This work however
establishes that the spin-down (or spin-up) time scale of a stably
stratified fluid is the classical Eddington-Sweet time scale, namely the
product of the heat diffusion time scale (also known as Kelvin-Helmholtz
time scale) and the factor $(\calN/2\Omega)^2$, where $\calN$ is the \BVF\
in the radiative region and $\Omega$ is the rotation rate. In slowly
rotating stars the Eddington-Sweet time scale is much longer than the
Kelvin-Helmholtz, but in fast rotators that are considered here, these
time scales are similar.  Hence, the spin-up driven by a gravitational
contraction has never been considered in the literature as far as we know.

The first step of our analysis considers the case of a neutrally
stratified envelope like the one met in a star with an outer
convection zone. The translation of this constant entropy medium into the
incompressible fluid model is the simple constant density fluid. This simple
model allows us to derive an asymptotic solution for small Ekman numbers
(i.e. small viscosity as appropriate for stellar applications). This
solution shows that the spinning up core controls the flow inside its
tangent cylinder and gives the amplitude of the quasi-steady flow
there. Outside the tangent cylinder we find that the solution depends
very much on the outer boundary conditions. We analyse the rather
artificial case where no-slip conditions are imposed on the outer
boundary, since this case also allows the derivation of an analytic
asymptotic solution. Moreover, with these boundary conditions, we can
also solve the case where the envelope is stably stratified (but not
contracting) and derive the associated differential rotation forced by
the baroclinic torque.

The competition between the two forcings (spin-up and baroclinicity)
necessarily arises in the pre-main sequence phase of an intermediate
mass star. We indeed expect that for such a mass range, an outer
radiative envelope sets in after the birth line, likely after the
disappearance of convective flows \cite[e.g.][]{maeder09}.

With our simplified model we can appreciate the result of this
competition. As a first step, still using the artificial no-slip
outer boundary condition, we compare the amplitudes of the
baroclinic flow and the contraction-driven spin-up flow when they are
taken separately.  It gives us a criterion on the parameters of the star.
Using numerical solutions of the full problem (including simultaneously
the two forcing mechanisms), we confirm the validity of the criterion.

Our next step is to leave aside the no-slip outer boundary condition and
concentrate on the more realistic stress-free condition. In such a case,
the derivation of an asymptotic analytic solution is much more involved
and we had to resort to numerics. Numerical solutions show that outside
the tangent cylinder of the core, a steady spin-up flow of an
unstratified fluid driven by contraction has an amplitude that scales
with the ratio of two small parameters, namely Ro/E, where Ro is the
Rossby number measuring the driving and E is the Ekman number measuring
the viscosity. With stellar parameters this ratio is expected to be
larger than unity showing that a steady solution is necessarily
of high amplitude. In addition, the time needed to reach such a steady
state is of the order of the contraction time, thus suggesting that this
steady state is actually never reached. This result prompted us to study
the transient state of the spin-up flow first without stratification and
then accounting for a stable stratification in the envelope. Before the
transient state reaches a significant amplitude, it can be studied with
linear equations that are easier to solve. Remarkably enough, this
transient flow shows the emergence of a quasi-self-similar solution that
simply grows in amplitude in the volume outside the core's tangent
cylinder. Since our original question is to know whether such flow is
able to supersede the baroclinic flow driven by the combination of
rotation and stratification, we compared the two flows. The easy way is
to compare the amplitudes of the flow taken separately and quite clearly
we find that the contraction-induced spin-up rapidly overtakes the amplitude of
a baroclinic flow. The numerical solution of the transient starting from
an established baroclinic flow confirms this result.

We now present the detailed derivation of these fluid dynamics results,
but the hurried reader more interested in the astrophysical side of the
problem may now jump to the discussion in section 5.

\subsection{Equations of motion}

In an inertial frame, the dynamics of an incompressible fluid enclosed
within the two shells is governed by the equations of momentum and mass
conservation :

\begin{equation}
\left \{
\begin{array}{ccl}
\frac{\partial \vec{v}}{\partial t} +(\vec{v}\cdot\vec{\nabla})\vec{v}
=-\frac{1}{\rho}\vec{\nabla}P + \nu \Delta\vec{v}\\
\vec{\nabla} \cdot \vec{v} =0
\label{eq1}
\end{array}
\right .
\end{equation}
where $\vec{v}$ is the velocity field, $P$ the pressure
and $\nu$ the kinematic viscosity of the envelope. Let us now remove the bulk
rotation and set

\begin{equation}
\vec{v}=\vec{\Omega} \wedge \vec{r} + \vec{w}\; .
\label{eq2}
\end{equation}
We define $\vO$ as the angular velocity of the core.
Since the field $\vec{v}$ is axisymmetric,

\[(\vec{v}\cdot\vec{\nabla})\vec{v}=\vec{\Omega} \wedge(\vec{\Omega}
\wedge \vec{r})+ 2\vec{\Omega} \wedge
\vec{w}+(\vec{w}\cdot\vec{\nabla})\vec{w} \; . \]
where we recognise the centrifugal and Coriolis accelerations.
The centrifugal potential is gathered with the pressure into
$\Pi$. Substituting (\ref{eq2}) in the set of equations (\ref{eq1}),
we find that the relative velocity $\vw$ verifies

\begin{equation}
\left \{
\begin{array}{ccl}
\frac{\partial \vec{w}}{\partial t}+ 2\vec{\Omega} \wedge
\vec{w}+(\vec{w}\cdot\vec{\nabla})\vec{w} = - \vec{\nabla}\Pi +
\nu\Delta\vec{w} - \vec{\dot\Omega} \wedge \vec{r}\\
\vec{\nabla} \cdot \vec{w} =0
\label{eq3}
\end{array}
\right .
\end{equation}
Let us note that the RHS-term now depends on the acceleration of the
rotation rate of the core $- \vec{\dot\Omega} \wedge \vec{r}$ where
$\vec{\dot\Omega}$ needs to be determined.

\subsection{Scaled equations and linearization}

We scale the equations using the Kelvin-Helmholtz time scale $\frac{R}{V_{s}}$, which is the
time scale of the complete absorption of the envelope by the core.  The length scale $R$
is the outer radius of the envelope, and the velocity scale is the suction
velocity $\vec{w}=V_{s}\vec{u}$.  The system (\ref{eq3}) now reads

\begin{equation}
\left \{
\begin{array}{ccl}
Ro\frac{D\vec{u}}{D\tau}+ \vec{e}_z \wedge \vec{u} = - \vec{\nabla}p +
E\Delta\vec{u} - \dot\omega \vec{e}_z \wedge \vec{r}\\
\vec{\nabla} \cdot \vec{u} =0
\label{eq4}
\end{array}
\right .
\end{equation}
where $p$ is the reduced pressure.
The following dimensionless numbers appear :

\begin{equation}
\dot\omega = \frac{\dot\Omega R}{2\Omega V_{s}},\quad
Ro=\frac{V_{s}}{2\Omega R},\quad E=\frac{\nu}{2\Omega R^2}
\end{equation}
Here we introduced the non-dimensional acceleration of the core rotation
rate $\dot\omega$, a Rossby number $Ro$ 
and the Ekman number $E$ which
measures the viscosity of the envelope.

Obviously, the suction velocity is very small compared to the rotation one.
Hence, we expect $Ro\ll1$. As a first step setting $Ro = 0$ seems reasonable as
long as $Ro\, u$ is less than unity (so as to be able to neglect quadratic
terms). Thus doing, we are left with a steady state problem which describes the
quasi-steady evolution of the system as long as the non-dimensional time $\tau$
verifies:

\begin{equation}
 Ro \ll \tau \ll \tau_s
 \label{eq13}
\end{equation}
where $\tau_s=\frac{\eta\tilde \rho}{3}$ is the scaled suction time
($\eta=\frac{R_{core}}{R}$ is the scaled inner radius). $\tau \gg Ro$
means that we neglect the transients corresponding to a few rotation
periods where boundary layers form. Likewise, $\tau\ll\tau_s$ means that
the rotation rate has not been changed, namely that $\dot\Omega
t\ll\Omega$.

\subsection{The acceleration of the core}

\subsubsection{General equation}

Equations (\ref{eq4}) need the expression of $\dot\omega$.  By absorbing
the envelope, the mass of the core grows, as its angular momentum.
Evolution of the angular momentum $L_{z}$ of the core is governed by

\begin{equation}
\frac{d{L_{z}}}{dt} = \vec{e}_z \cdot \{-\int_{(S)}(\vec{r} \wedge
\rho\vec{v})\vec{v} \cdot \vec{dS}+\int_{(S)}\vec{r} \wedge
[\sigma]\vec{dS}\}
\end{equation}
The first integral is the flux of incoming angular momentum and the
second one is the viscous torque applied on the core surface.
$[\sigma]$ is the stress tensor. We have

\begin{equation}
\vec{e}_z \cdot (\vec{r} \wedge
[\sigma]\vec{e_r})=r\sin\theta\sigma_{r\phi}=r\sin\theta \rho\nu
\left.\frac{\partial w_\phi}{\partial r}\right|_{r=R_{core}}
\end{equation}
Besides, for a sphere of mass $M_{core}$ and radius $\eta R$ :

\begin{equation}
\vec{L}=I\vec{\Omega}=\frac{2}{5}M_{core}(\eta R)^2\vec{\Omega}
\end{equation}
With the previous scaling, we get

\begin{equation}
\begin{array}{ccl}
M_{core} \dot\Omega &=& \frac{8\pi}{3}\Omega \rho V_{s} (\eta
R)^2 \\
 &+& 5\pi \nu \rho V_{s} \eta \displaystyle\int_0^\pi \sin^2\theta
\left.\frac{\partial u_\phi}{\partial r} \right|_{r=\eta} d\theta\\
\label{eq6}
\end{array}
\end{equation}
The evaluation of the remaining integral needs the expression of the
azimuthal flow $u_{\phi}$ at the core-envelope boundary $r=\eta$, namely in the Ekman boundary layers.

\subsubsection{Boundary layer analysis}

First of all, we change the boundary conditions with mass-flux to ordinary
boundary condition by making the substitution

\[ \vec{u}=\vec{u'} - \frac{\eta^2}{r^2} \vec{e}_r\]
The new velocity field $\vec{u'}$ verifies

\begin{equation}
\left \{
\begin{array}{ccl}
\vec{\nabla}\wedge\{\vec{e}_z \wedge \vec{u'} - E\Delta\vec{u'}\}&=&2(
\frac{\eta^2}{r^3} - \dot\omega) \cos\theta \vec{e}_r 
\\&+&(\frac{\eta^2}{r^3} + 2\dot\omega) \sin\theta \vec{e}_\theta\\
\vec{\nabla} \cdot \vec{u'} &=&0
\label{eq5}
\end{array}
\right .
\end{equation}
with the boundary conditions

\[ \vu' = \vzero \on r=\eta\]
and

\[ \er\times[\sigma]\er=\vzero, \andet u'_r=0 \on r=1\]
These latter conditions describe stress-free boundary conditions. Indeed,
we assume that the upper layers outside the outer shell have unimportant
dynamic effects and are just providing/absorbing some mass. In the
following, we drop the prime on the velocity field.

As shown in \cite{ELR13}, the Ekman number in stars is always very
small thus leading to the formation of boundary layers near the boundaries.
To derive the expression of the flow, we therefore examine the asymptotic
case $E \ll 1$. We first note that if we consider the inviscid case $E=0$,
the set of equations (\ref{eq5}) is solved by

\begin{equation}
\left \{
\begin{array}{ccl}
\bar{u}_r&=&\frac{\eta^2}{r^2}+r\dot
\omega\left(2-3\sin^2\theta\right)\\
\bar{u}_\theta&=&-3r\dot \omega\sin\theta\cos\theta\\
\end{array}
\right .
\end{equation}
In the azimuthal direction, we look for a flow such as
$\bar{u}_\phi=U(s)\vec{e_{\phi}}$ as dictated by the Taylor-Proudman
theorem \cite[e.g.][]{Green69}. Such a flow does not verify the no-slip
boundary conditions at the interface $r=\eta$. It needs boundary layer
corrections so as to satisfy $\bar{u}_0+\widetilde{u}_0=0$.  The bar
refers to the solution within the envelope i.e. outside the boundary
layer and tilded quantities are for boundary layer corrections.

Following \cite{R06} we write the boundary layer corrections as:

\begin{equation}
(\vec{n}\wedge\widetilde{u}_0+i\widetilde{u}_0)=-(\vec{n}\wedge\bar{u}_0+i\bar{u}_0)_{\alpha=0}\exp{(-(1-i)\alpha)}
\end{equation}
where $\displaystyle{\alpha=(r-\eta)\sqrt{\frac{|\cos\theta|}{2E}}}$
and $\vec{n}=-\vec{e}_r$. We only keep the decreasing solution.  
The corrections thus read

\begin{equation}
\left \{
\begin{array}{ccl}
\widetilde{u}_\theta&=&\displaystyle{-U(\eta \sin  \theta)\ \sin
\alpha\ e^{-\alpha}}\\
\widetilde{u}_\phi&=&\displaystyle{-U(\eta \sin  \theta)\  \cos  \alpha\
e^{-\alpha}}
\end{array}
\right .
\end{equation}
As in \cite{R06}, mass conservation gives the relation between the Ekman pumping
$\widetilde{u}_r$ and  the geostrophic flow $U$.
It yields

\begin{equation}
1 + (2-3\sin^2\theta)\eta\dot \omega=\frac{1}{\eta \sin
\theta}\sqrt{\frac{E}{2}}\frac{\partial}{\partial \theta} \left
(\frac{\sin  \theta\ U(\eta \sin  \theta)}{\sqrt{|\cos  \theta|}} \right
)
\end{equation}
where we keep only $\mathcal{O}(\sqrt{E})$ terms.
Integrating with respect to $\theta$, we get 

\begin{equation}
U(\eta
\sin\theta)=\eta\sqrt{\frac{2}{E}}\frac{{\sqrt{|\cos\theta|}}}{\sin\theta}\left(
1-\cos\theta + \eta \dot \omega \cos\theta \sin^2\theta \right) \label{Usol}
\end{equation}
Note that this expression defines $U$ only within the tangent cylinder of
the core defined as $s=r\sth =\eta$ ($s$ is the radial cylindrical
coordinate).

Near the surface $r=\eta$ i.e. within the boundary layer, the shear is
dominated by the boundary layer correction.  It simplifies the computation
of the radial derivative of the azimuthal flow $u_\phi$ which reads

\begin{equation}
\frac{\partial u_\phi}{\partial r}=\frac{\eta}{E}|\cot\theta|\left(
1-\cos\theta + \eta \dot \omega \cos\theta \sin^2\theta \right)
\end{equation}
Integral in \eq{eq6} can now be evaluated:

\begin{equation}
\int_0^\pi \sin^2\theta
\left.\frac{\partial u_\phi}{\partial r}\right|_{r=\eta} d\theta =
\eta \frac{2}{3E}\left( \frac{1}{2} + \frac{2}{5} \eta \dot \omega
\right)
\end{equation}
We can now derive the acceleration of the angular velocity of the core.
Considering quasi-steady solutions that arise when $Ro \ll \tau \ll \tau_s $, \eq{eq6} leads to

\begin{equation}
\dot\omega = \frac{9}{4\tilde\rho \eta}
\label{eq18}
\end{equation}
which completes the equations (\ref{eq5}).

The foregoing solution \eq{Usol} shows that the differential rotation
driven by the mass contraction scales as $\mathcal{O}(E^{-1/2})$. It means
that the linear regime that we solved is valid only when $Ro\ll \sqrt{E}$,
which is actually the case (see below).  The foregoing solution
however describes the fluid flow only within the tangent cylinder of the
core.

Outside the cylinder, the solution depends on the outer boundary
conditions and on the Stewartson layer that lies upon the cylinder. This
makes the global solution quite involved, all the more that we should also
account for a possible stable stratification of the envelope.  Indeed, during
the PMS phase of intermediate mass stars, the envelope is completely
radiative. Therefore the contraction-induced differential rotation
competes with the one induced by the baroclinicity of the envelope.

Before getting any further, we need evaluating stellar numbers that
have appeared.

\subsection{Orders of magnitude}

As a test case of the foregoing problem, we first consider the contraction
of a fully radiative 3~\msun\ PMS star. On the birthline, its surface
temperature is around $T_*\sim5600K$, its luminosity $L_*\sim10^2L_\odot$
and its radius is $R_*\sim10^{10}\text{m}$.  The young star contracts
on the Kelvin-Helmholtz time upon the PMS (Henyey) track, namely:

\[ t_{\rm KH} = \frac{GM^2}{RL}\]
according to \cite{maeder09}. This leads to $t_{\rm
KH}\sim2.6~10^5\text{yr}$.

Setting arbitrarily $R_{\rm core}=0.15R_*$, we find $V_{s}\sim 6\times
10^{-5}\text{m.s$^{-1}$}$. Considering a rotation velocity of 10~km/s (such that
near the end of the PMS after a gravitational contraction at constant angular
momentum we obtain a star like
HD 37806 studied by \citealt{BC95}) we find
a Rossby number

\[ Ro\sim 3\times 10^{-9}\]
which is very small as expected.

The estimate of the Ekman number requires a value of the kinematic
viscosity.  If we use Zahn's prescription \cite[][]{zahn92} for a
turbulent viscosity, we get $\nu\sim10^{4}\text{ m$^2\cdot$s$^{-1}$}$ and thus

$$E\sim 10^{-10}$$
With the radiative viscosity \cite[e.g.][]{ELR13}, $\nu\sim10^{2}\text{
m$^2\cdot$s$^{-1}$}$, we get

$$E\sim 10^{-12}$$
For both values the condition $Ro \ll \sqrt{E}$ is satisfied.

During the contraction, the quasi-steady state within
the cylinder is reached after a spin-up time, namely after
$\frac{(2\Omega)^{-1}}{\sqrt{E}}$. Using the previous numbers, we find
that this state occurs after $\sim10^3$ or $10^4$ years so rather shortly
after the start of contraction. Therefore, within the tangent cylinder,
we can neglect the transient phase.

If we consider the contracting envelope of a Jupiter-like giant planet,
Ekman numbers are also very small although with larger uncertainties:
\cite{OL04} give $E\sim10^{-7}$ while \cite{wu05} suggest
$E\sim10^{-13}$. The typical contraction time of giant planets is over
1~Gyrs \cite[][]{fortney_etal10} so that $Ro \sim P_{\rm rot}/t_{\rm
contraction} \sim 10^{-12}$. The condition $Ro \ll \sqrt{E}$ is also
easily satisfied.

\subsection{Adding stratification}

\subsubsection{Scaled equations}

To account for a stable stratification in the envelope we now generalize
the set of equations (\ref{eq5}) by taking into account the buoyancy
force and the equation for temperature fluctuations. In PMS stars the
stable stratification of the envelope may come after a convective episode
and thus may be evolving with time. To simplify and get an upper bound on
the effects of stratification we take the \BVF\ as constant in time.

To be consistent with the foregoing model that uses an incompressible
fluid we use the Boussinesq approximation. Following \cite{R06} and
combining with (\ref{eq5}) we find

\begin{equation}
\left \{
\begin{array}{l}
\vec{\nabla} \times (\vec{e}_z \wedge \vec{u} -B \theta_T \vec{r} -
E\Delta\vec{u})= -B n_T^2 \sin{\theta}\cos{\theta}\vec{e}_{\phi} \\
-2( \frac{\eta^2}{r^3} + \dot\omega) \cos\theta \vec{e}_r
 +(\frac{\eta^2}{r^3} +
2\dot\omega)\sin\theta \vec{e}_\theta \\
\\
(n_T^2 / r)u_r =B\tilde{E}_T\Delta\theta_T\\
\\
\vec{\nabla}\cdot\vec{u}=0
\end{array}
\right .
\label{eq7}
\end{equation}
We use the same scales and notations as \cite{R06}.  The temperature
perturbation is scaled as $\delta T = \epsilon T_* \theta_T$ where
$\epsilon=\frac{\Omega^2R}{g_s}$ is the ratio of centrifugal acceleration
to surface gravity. Recall that the centrifugal acceleration is driving
the baroclinic flow.  The Brunt-V\"ais\"al\"a profile $n_T^2$ is
scaled with $\mathcal{N}^2=\frac{\alpha T_*g_S}{R}$ where $\alpha$
is the dilation coefficient and $g_s$ the surface gravity.

\begin{figure}
\begin{center}
\includegraphics[width=0.5\textwidth,height=0.25\textheight,angle=0]{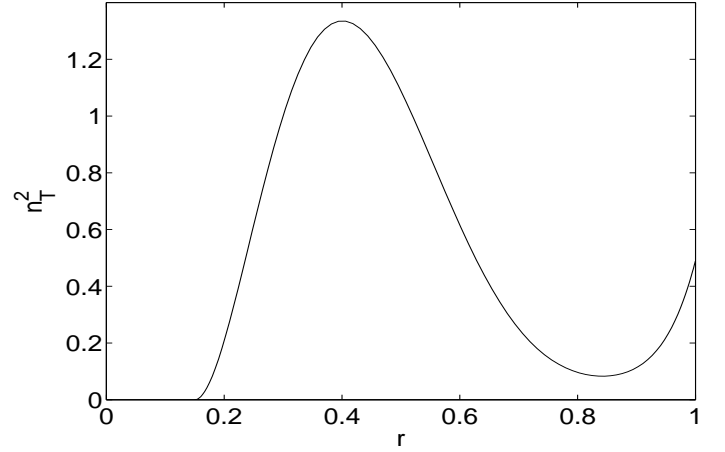}
\caption{The typical and scaled Brunt-V\"ais\"al\"a frequency profile as
a function of the normalized radius with $\eta=0.15$.\label{fig2}}
\end{center}
\end{figure}

The dimensionless number $B$ monitors the ratio of the forcings. From the
expression of the scaling of the baroclinic flow \cite[see][]{R06}, we
have

\beq B=\frac{\epsilon \mathcal{N}^2 R}{2\Omega V_{s}} \eeq
Finally, the dimensionless number

\beq \tilde{E}_T=\frac{E}{\lambda}\qquad {\rm with}\quad \lambda =
\PR\frac{\calN^2}{4\Omega^2}\eeq
measures heat diffusion. $\PR$ is the Prandtl number. For fast
rotators, $\lambda$ is a small parameter which we set to $10^{-4}$
following the estimate of \cite{RB14}.

Based on stellar models, a typical profile of the Brunt-V\"ais\"al\"a
frequency is shown in Fig.~\ref{fig2}. We use the polynomial expression

\begin{equation}
\left \{
\begin{array}{ccl}
n_T^2(r)  &=& 0  \text{ if }r< \eta \\
\\
n_T^2(r) &=& (\alpha (r-\eta) + \beta (r-\eta)^2 + \gamma(r-\eta)^3)^2
\text{ if }r  \in[\eta;1].
\end{array}
\right.
\end{equation}
to represent this function. $\alpha$, $\beta$ and $\gamma$ are the coefficients resulting from
the polynomial fit.

\section{An interesting solution with rigid outer boundary conditions}

Before getting into the full numerics, it is interesting to consider
the case where the outer bounding sphere of the envelope is rigidly
rotating at the same rate as the inner core. Outer no-slip conditions
can be expected if a turbulent layer threaded by magnetic fields covers
the stellar surface \cite[see][]{RB14}, however the synchronism between
this layer and the surface is here an {\it ad hoc} assumption (which can be
easily removed actually). The interesting point that is addressed below
comes from the simple analytical solutions that can be derived for the
flow outside the tangent cylinder and which gives an interesting view of
the properties of the system.

\subsection{The steady mass contraction induced flow}

\begin{figure}
 	\begin{center}
\includegraphics[width=0.5\textwidth,angle=0]{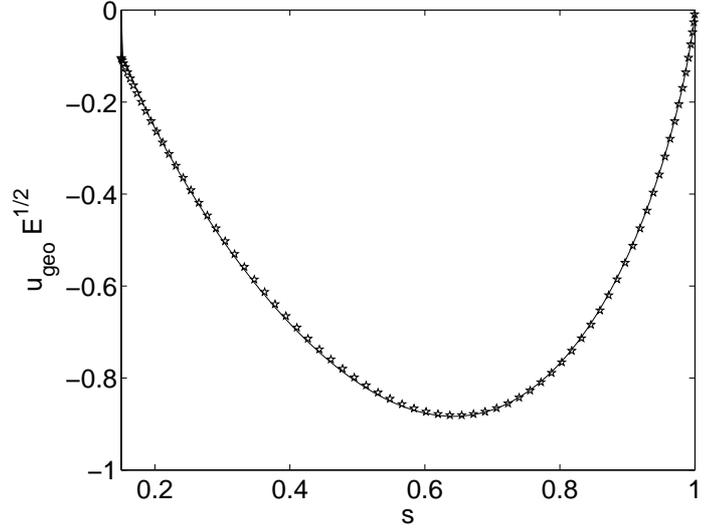}
 	\end{center}
\caption{Comparison between numerical (solid line) and analytical
(star line) solutions of the geostrophic flow at the equator
$u_\phi(r,\theta=\frac{\pi}{2})$ for $E=10^{-7}$, $\eta=0.15$, $\tilde
\rho=10$ without stratification when the envelope is rigidly rotating at the
same rate as the inner one.  \label{fig3}}
\end{figure}

With no-slip conditions on the outer boundary $r=1$, we may easily derive
the expression of the geostrophic flow\footnote{A geostrophic flow is a
steady flow that realizes the perfect balance between the Coriolis force
and the pressure gradient. As a consequence, it does not depend on the
coordinate parallel to the rotation axis (Taylor-Proudman theorem) and
behaves as a columnar flow.} out of the tangent cylinder of
the core. When no stratification is present, the azimuthal velocity reads

\begin{equation}
U(s)=\sqrt{\frac{2}{E}}(1-s^2)^{3/4}\left( \frac{\eta^2}{s} -\dot \omega
s \right) \qquad s\geq\eta
\label{eq12}
\end{equation}
As shown in Fig.~\ref{fig3}, this analytical solution nicely matches the
numerical one.

\subsection{The steady baroclinic flow}

\begin{figure}
 \begin{center}
 \includegraphics[width=0.5\textwidth,angle=0]{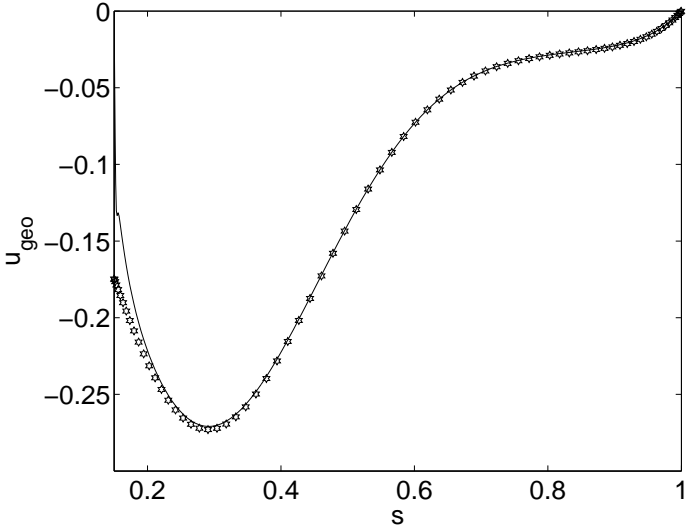}
 \end{center}
\caption{Comparison between numerical (solid line) and analytical
(stars) solutions of the geostrophic flow at the equator
$u_\phi(r,\theta=\frac{\pi}{2})$ 
for $E=10^{-7}$, $\eta=0.15$, $\tilde \rho=10$, $B=1$ and $\lambda=10^{-4}$
without mass contraction.\label{fig4}}
\end{figure}

Let us now consider the opposite case where a pure baroclinic flow
(no contraction) meets no-slip boundary conditions. It verifies

\begin{equation}
\left \{
\begin{array}{ccl}
\vec{\nabla} \times (\vec{e}_z \wedge \vec{u} -B \theta_T \vec{r} -
E\Delta\vec{u})=-B n_T^2 \sin{\theta}\cos{\theta}\vec{e}_{\phi}\\
(n_T^2 / r)u_r =B\tilde{E}_T\Delta\theta_T\\
\vec{\nabla}\cdot\vec{u}=0
\label{eq8}
\end{array}
\right .
\end{equation}
As shown in \cite{R06}, neglecting temperature perturbations and viscosity, the $\phi-$component of the vorticity equation (\ref{eq8})
leads to

\begin{equation}
\left \{
\begin{array}{lcl}
u_\phi &=& - s B \int_{r}^{1} \frac{n^2(r)}{r} dr + F(s)\\
\theta_T &=&0
\label{eq9}
\end{array}
\right .
\end{equation}
where $F(s)$ is a geostrophic solution determined by the boundary conditions.
In order to get the expression of $F(s)$, we write

\begin{equation}
\bar{u}_0=\left(- s B \underbrace{\int_{r}^{1} \frac{n^2(r)}{r}
dr}_{g(r)} + F(s) \right) \vec{e}_\phi
\end{equation}
and look for the boundary layer corrections at $r=1$. These are

\[\widetilde{u}_0=\mathcal{I}m\{-(\vec{e}_r
\wedge\bar{u}_0+i\bar{u}_0)_{r=1}e^{-\alpha(1+i)}\}\]
where $\displaystyle{\alpha=(1-r)\sqrt{\frac{|\cos\theta|}{2E}}}=\zeta\sqrt{\frac{|\cos\theta|}{2}}$. It yields

\begin{equation}
\left \{
\begin{array}{ccl}
\widetilde{u}_\theta=-(-\sin\theta B g(1)+F(\sin\theta))\sin\alpha
e^{-\alpha}\\
\widetilde{u}_\phi=-(-\sin\theta B g(1)+F(\sin\theta))\cos\alpha
e^{-\alpha}
\end{array}
\right .
\end{equation}
We note that $g(1)=0$. Mass conservation implies
\begin{equation}
 \frac{1}{\sqrt{E}}\frac{\partial \widetilde{u}_r}{\partial \zeta}=\frac{1}{\sin\theta}\frac{\partial (\sin\theta \widetilde{u}_\theta)}{\partial \theta}
\end{equation}
so that

\begin{equation}
\widetilde{u}_r(1)
=-\frac{\sqrt{E}}{\sth}\dtheta{}\lp\sth\int_\zeta^\infty\widetilde{u}_\theta
d\zeta\rp\; .
\end{equation}
Since $\tu_r+\bu_r=0$ at $r=1$, we finally get

\begin{equation}
 -\bar{u}_r(r=1)=\sqrt{\frac{E}{2}}\frac{1}{\sin\theta}\frac{\partial}{\partial \theta} \left(\frac{\sin\theta}{\sqrt{\cos\theta}} F(\sin\theta)\right)
\; .\label{bcbur}
\end{equation}
However, the radial component of the vorticity equation (or the angular momentum
equation) in the interior leads to

\[ \sth \bu_r+\cth \bu_\theta = E \Delta'\bu_\varphi\; .\]
Here $\bu_\varphi=-s B g(r)+F(s)$ so that consistency of the solution with
\eq{bcbur} requires that

\[ F \equiv \od{B\sqrt{E}}\; .\]
It means that in the limit of vanishing Ekman numbers the function $F$
can be neglected. Therefore, at leading order, the envelope differential
rotation is dominated by the shellular flow :

\[ \bar{u}_\phi = - s B \int_{r}^{1} \frac{n^2(x)}{x} dx = -s B g(r)\; .\]
Fig.~\ref{fig4} shows the comparison of the analytical solution $-s
B g(r)$ with the numerical one at the equator $\theta=\frac{\pi}{2}$.
The difference on the left edge comes from the fact that $g(\eta)$
is not zero and would require an additional boundary layer correction.

\begin{figure}[h]
 	\begin{center}
 \includegraphics[width=0.3\textwidth,height=0.135\textheight,angle=0]{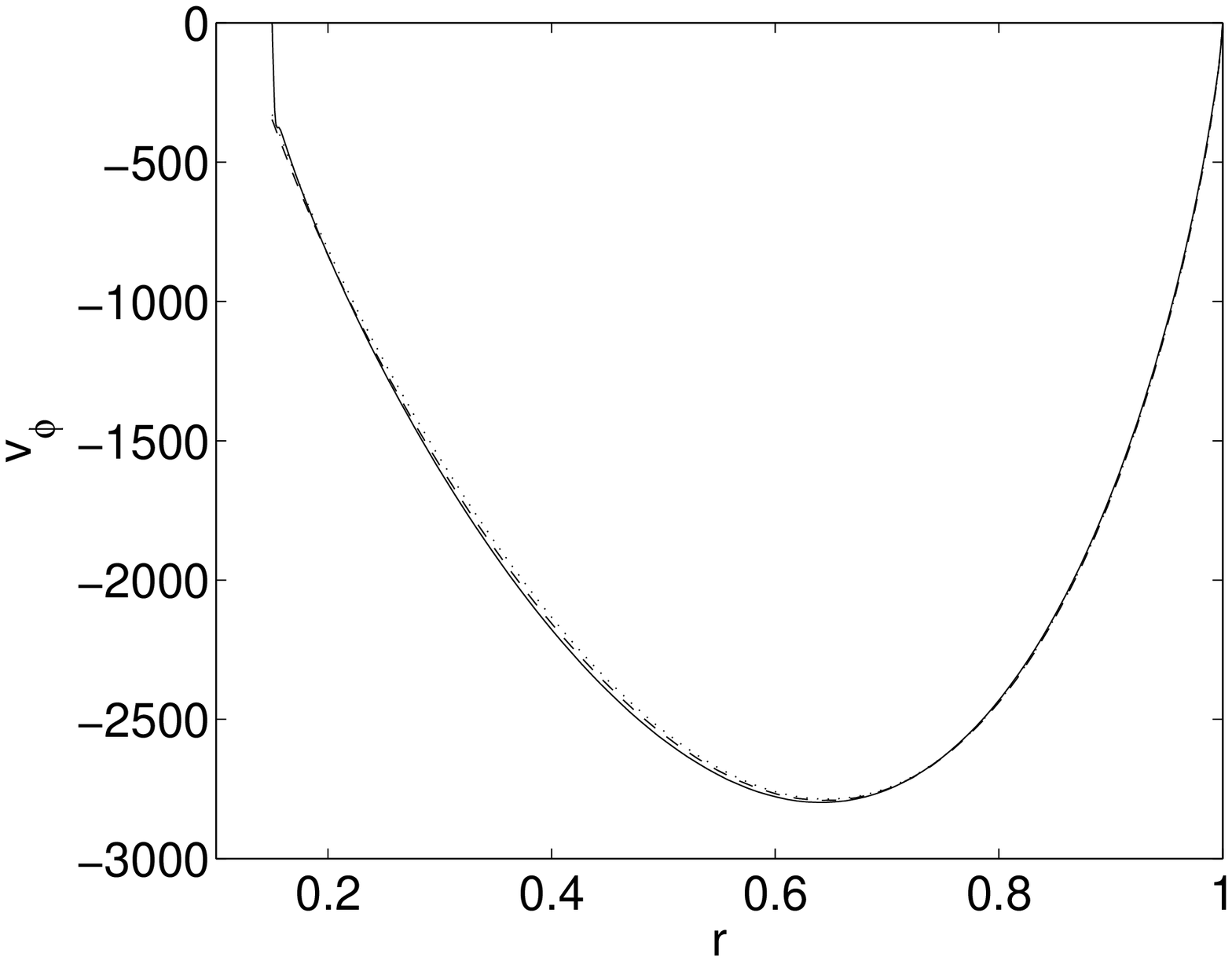}\\	
 \includegraphics[width=0.3\textwidth,height=0.135\textheight,angle=0]{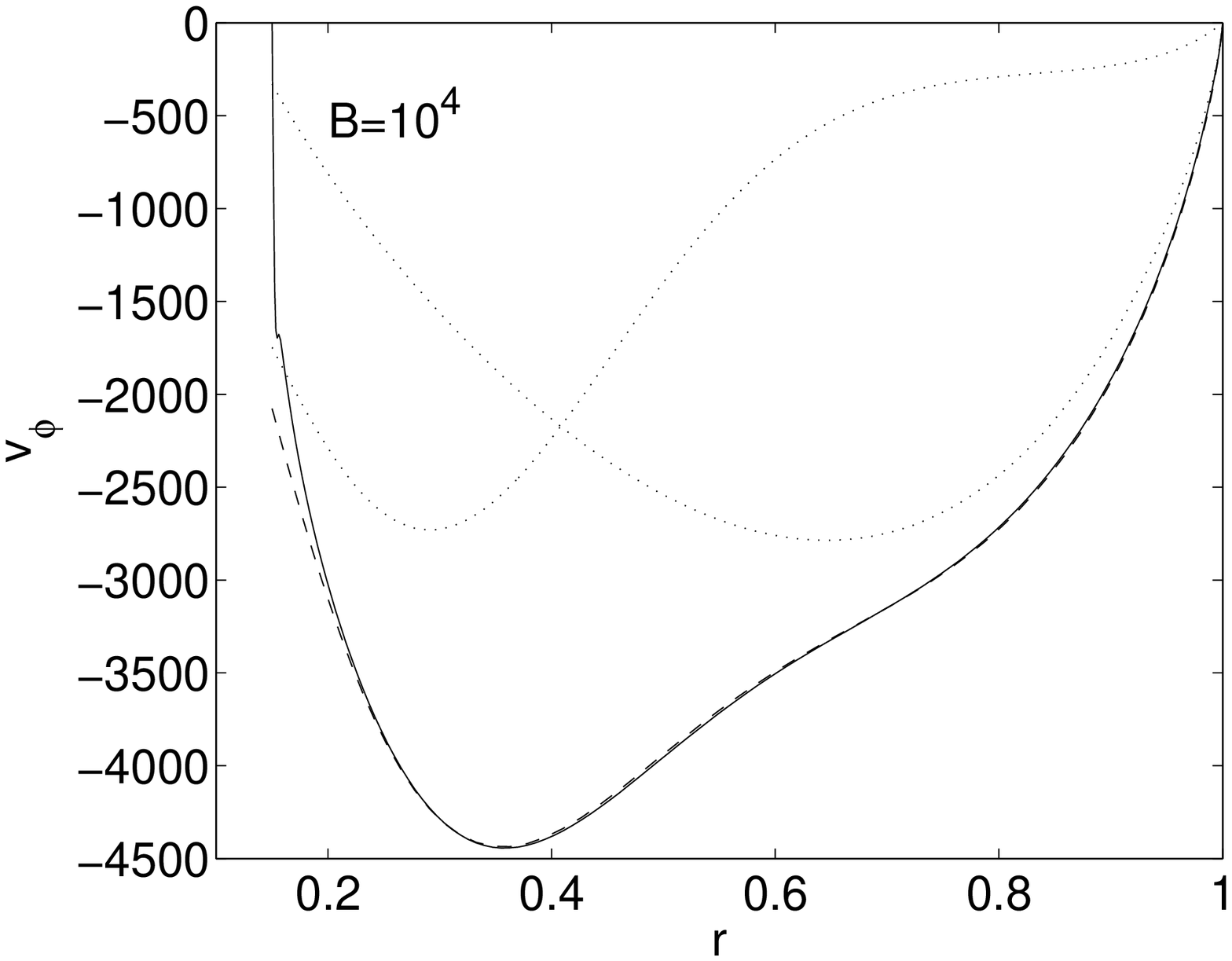}\\	
 \includegraphics[width=0.3\textwidth,height=0.135\textheight,angle=0]{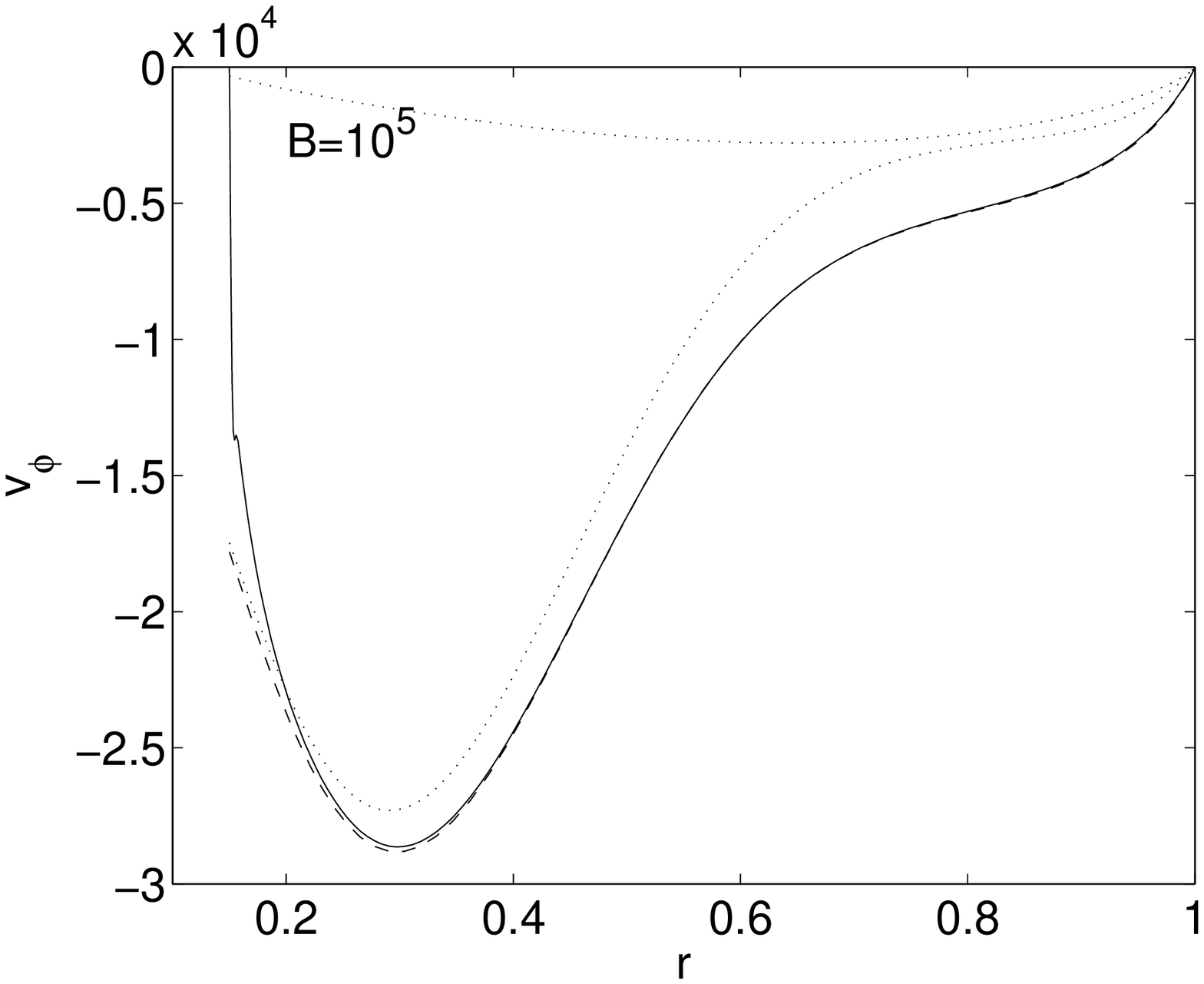}\\
 \includegraphics[width=0.3\textwidth,height=0.135\textheight,angle=0]{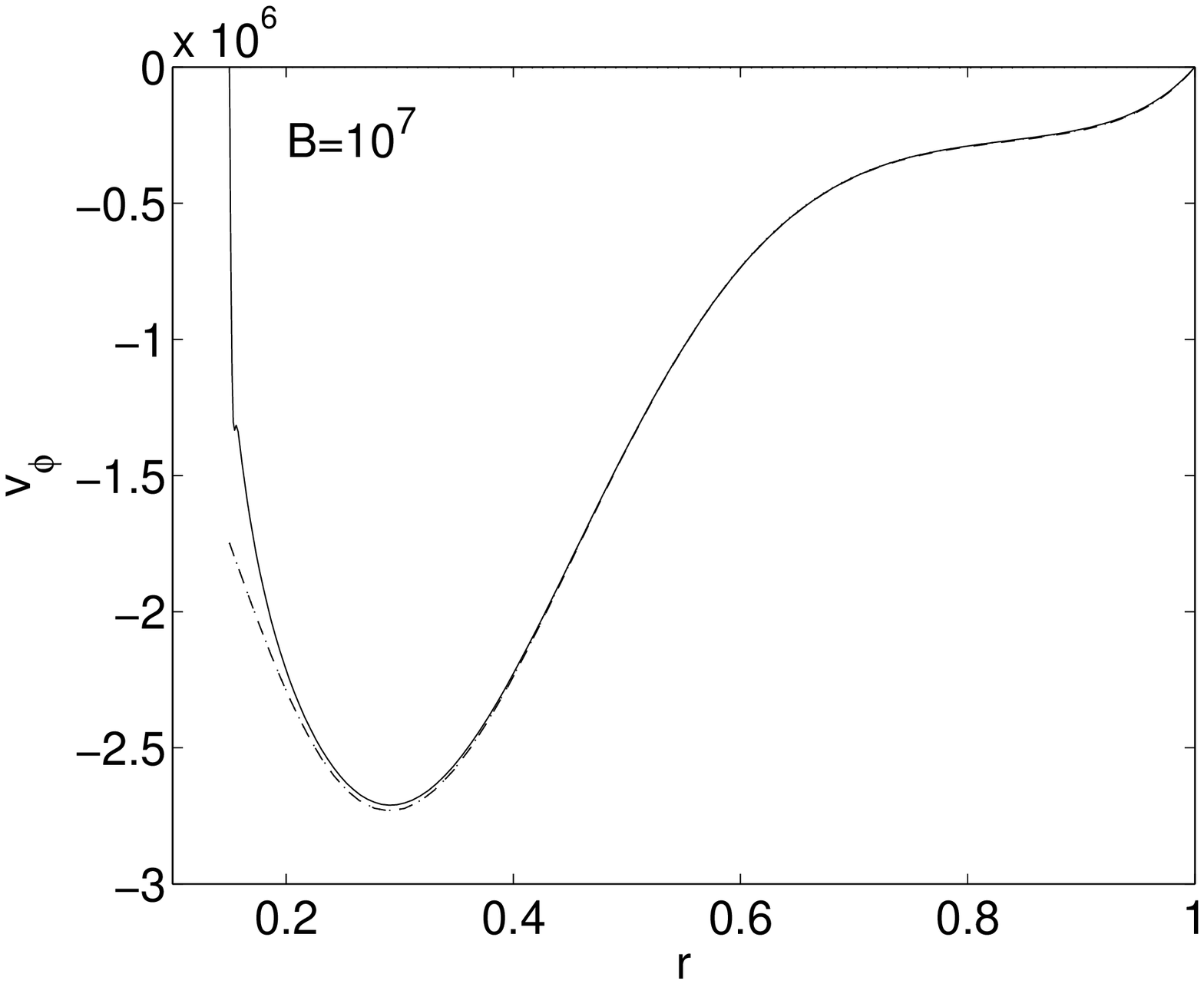}\\
 \end{center}
\caption{Evolution of the azimuthal velocity
 $u_{\phi}(r,\theta=\frac{\pi}{2})$ for $\eta=0.15$, $\tilde\rho=10$,
 $E=10^{-7}$, $\lambda=10^{-4}$ and $B=0,10^4,10^5,10^7$ (downward).
 The solid line is the numerical solution, the
 dashed line is the full analytical solution and the two dot lines are
 the solutions of each flow. The transition is at $B=10^4$, the
 differential rotation is then governed by baroclinicity as shown in Fig.~\ref{fig5}.
 Boundary conditions are no-slip on both
 sides $r=\eta$ and $r=1$. \label{fig1}}
\end{figure}

\subsection{Transition between the two steady flows : the baroclinic
flow versus the one induced by mass contraction}

\begin{figure*}
 	\begin{center}
 \includegraphics[width=0.3\textwidth,height=0.216\textheight,angle=0]{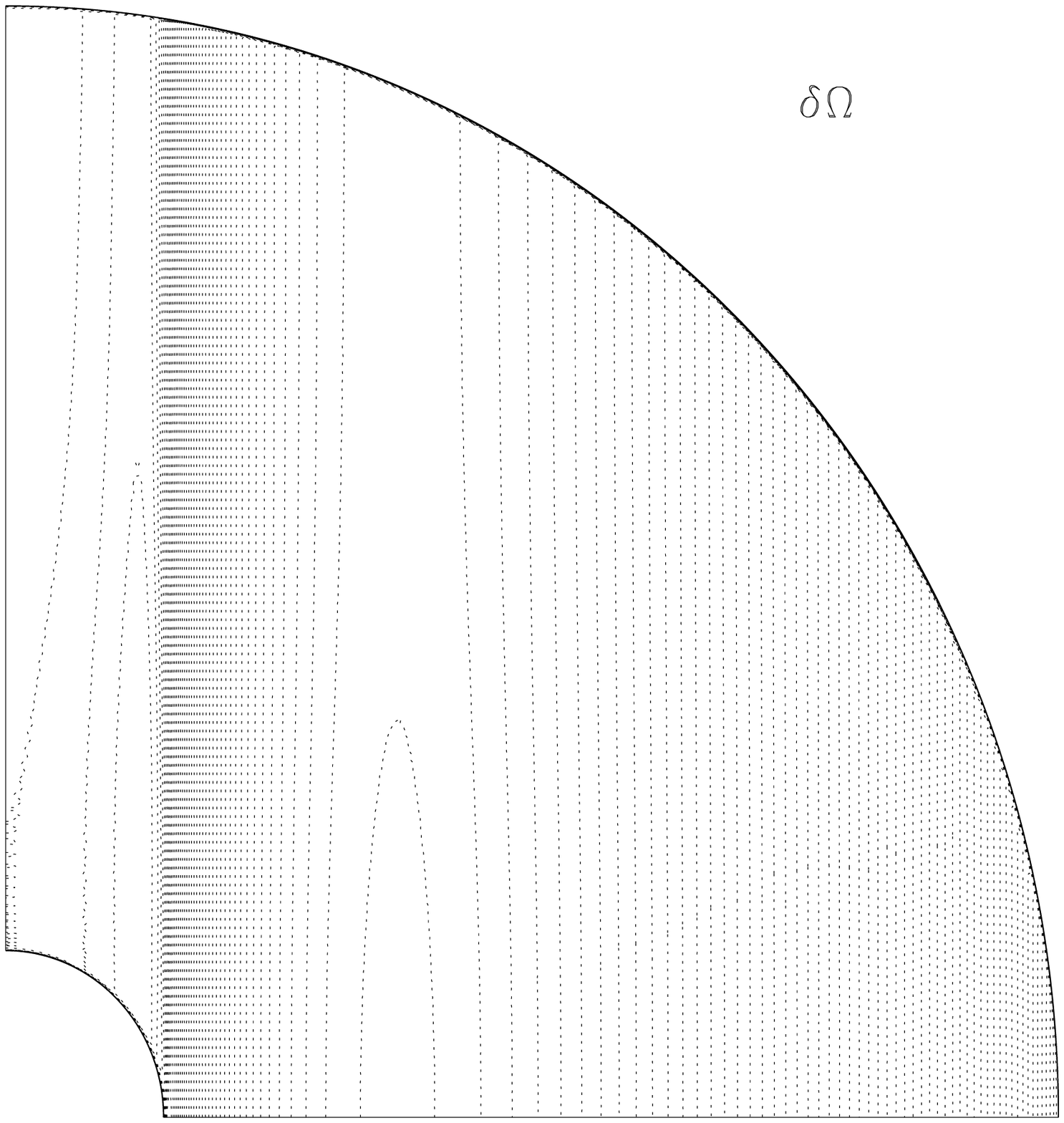}
 \includegraphics[width=0.3\textwidth,height=0.216\textheight,angle=0]{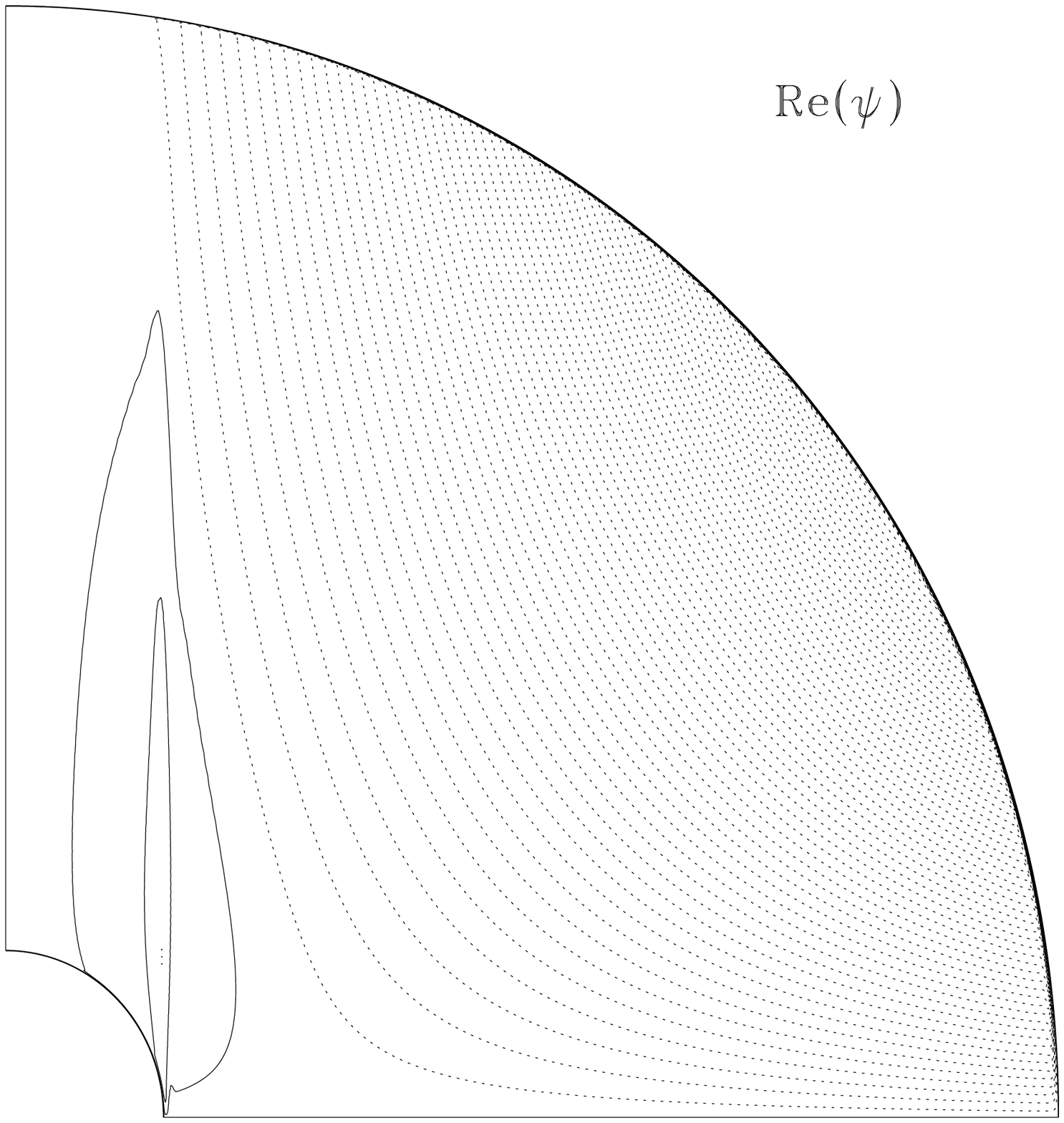}\\
 \includegraphics[width=0.3\textwidth,height=0.216\textheight,angle=0]{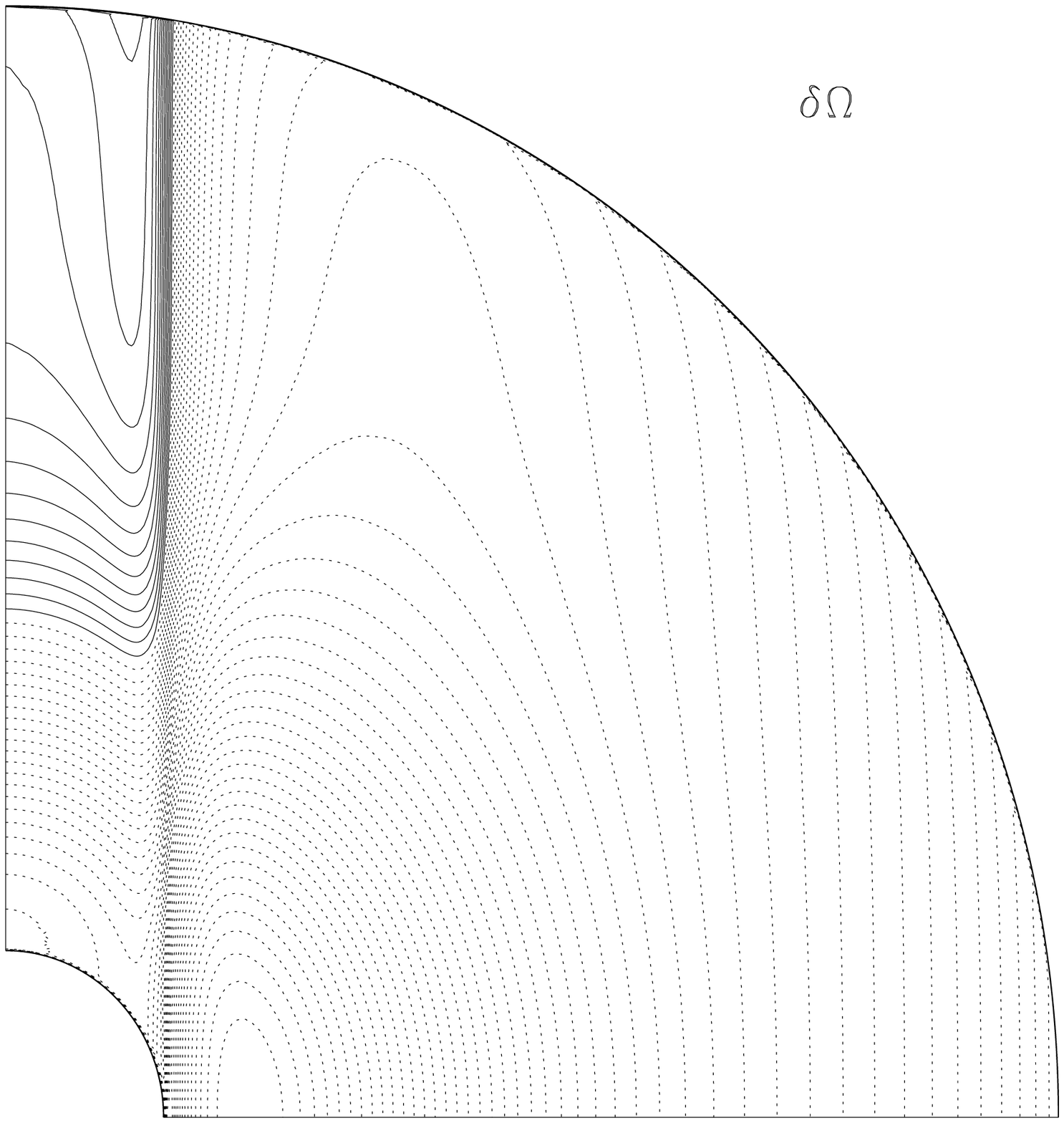}
 \includegraphics[width=0.3\textwidth,height=0.216\textheight,angle=0]{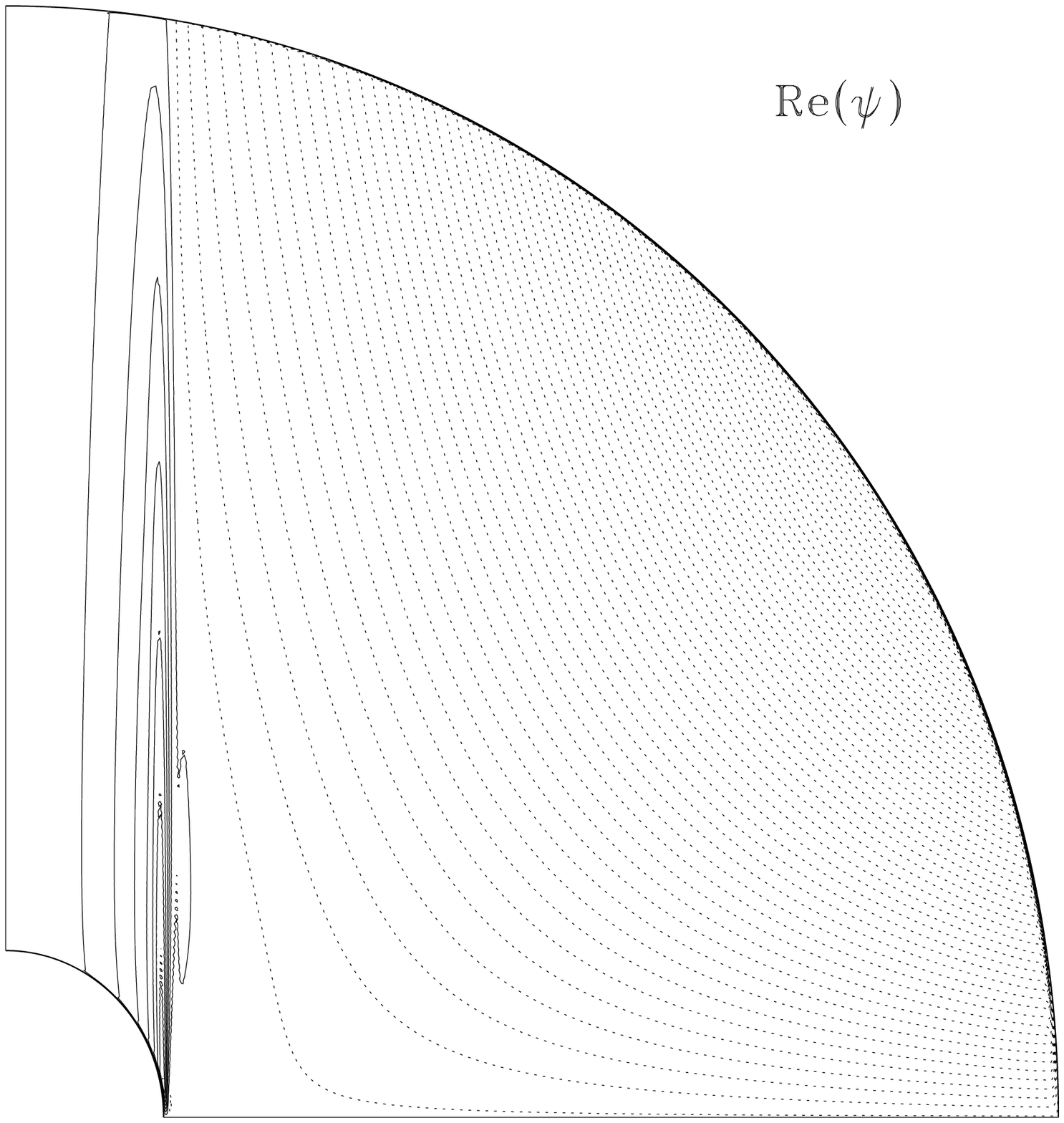}\\
 \includegraphics[width=0.3\textwidth,height=0.216\textheight,angle=0]{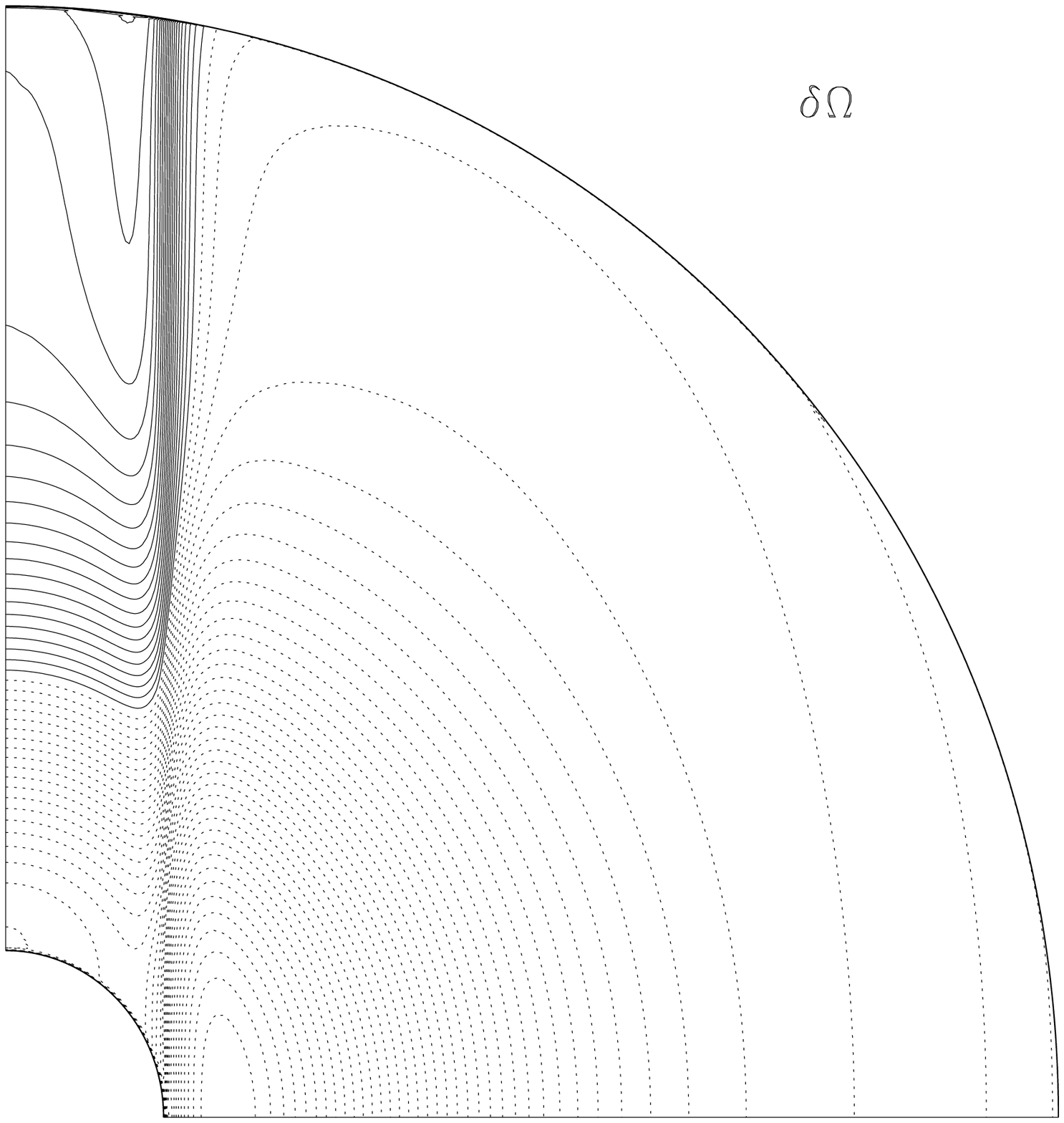}
 \includegraphics[width=0.3\textwidth,height=0.216\textheight,angle=0]{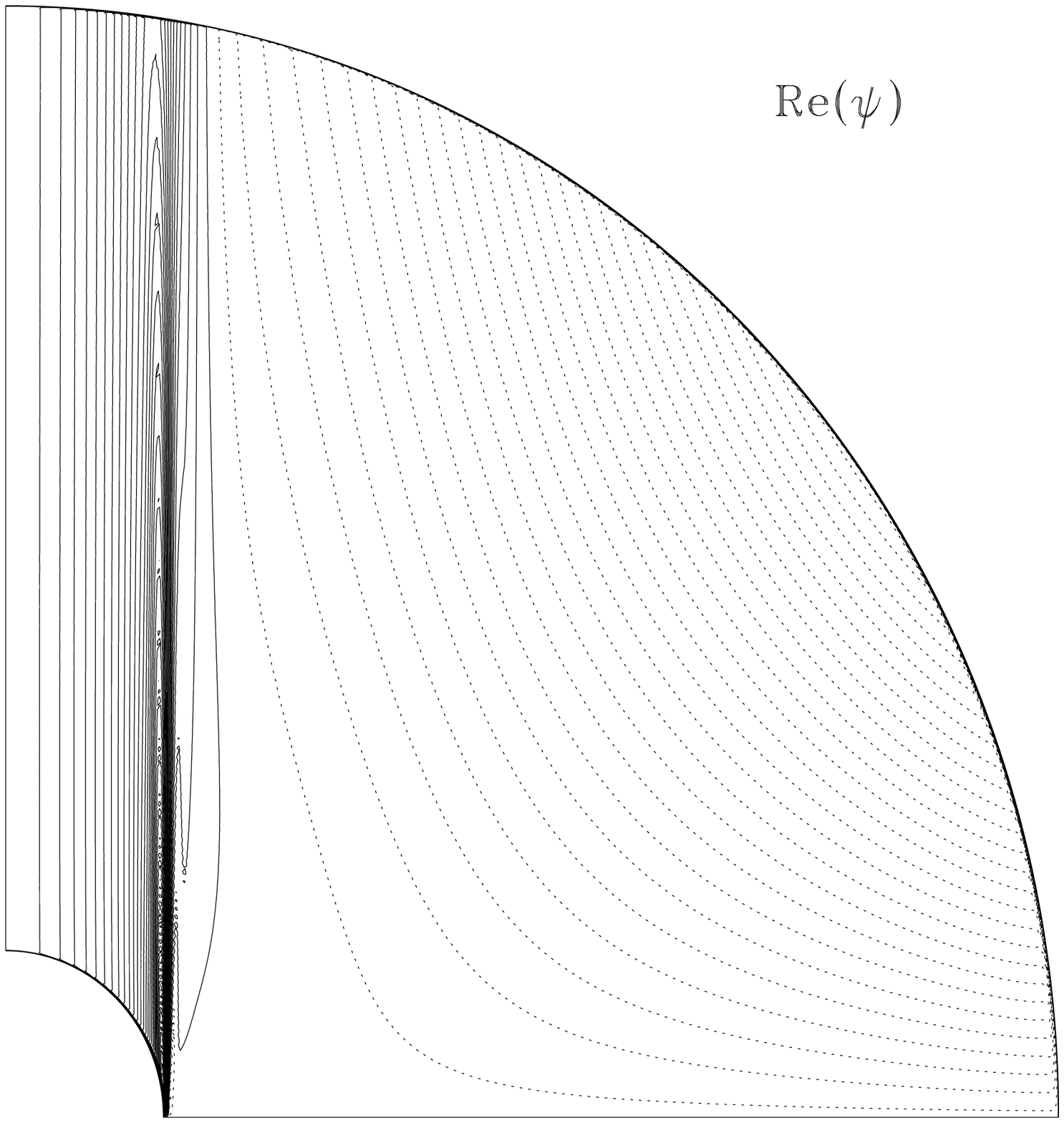}\\
 \includegraphics[width=0.3\textwidth,height=0.216\textheight,angle=0]{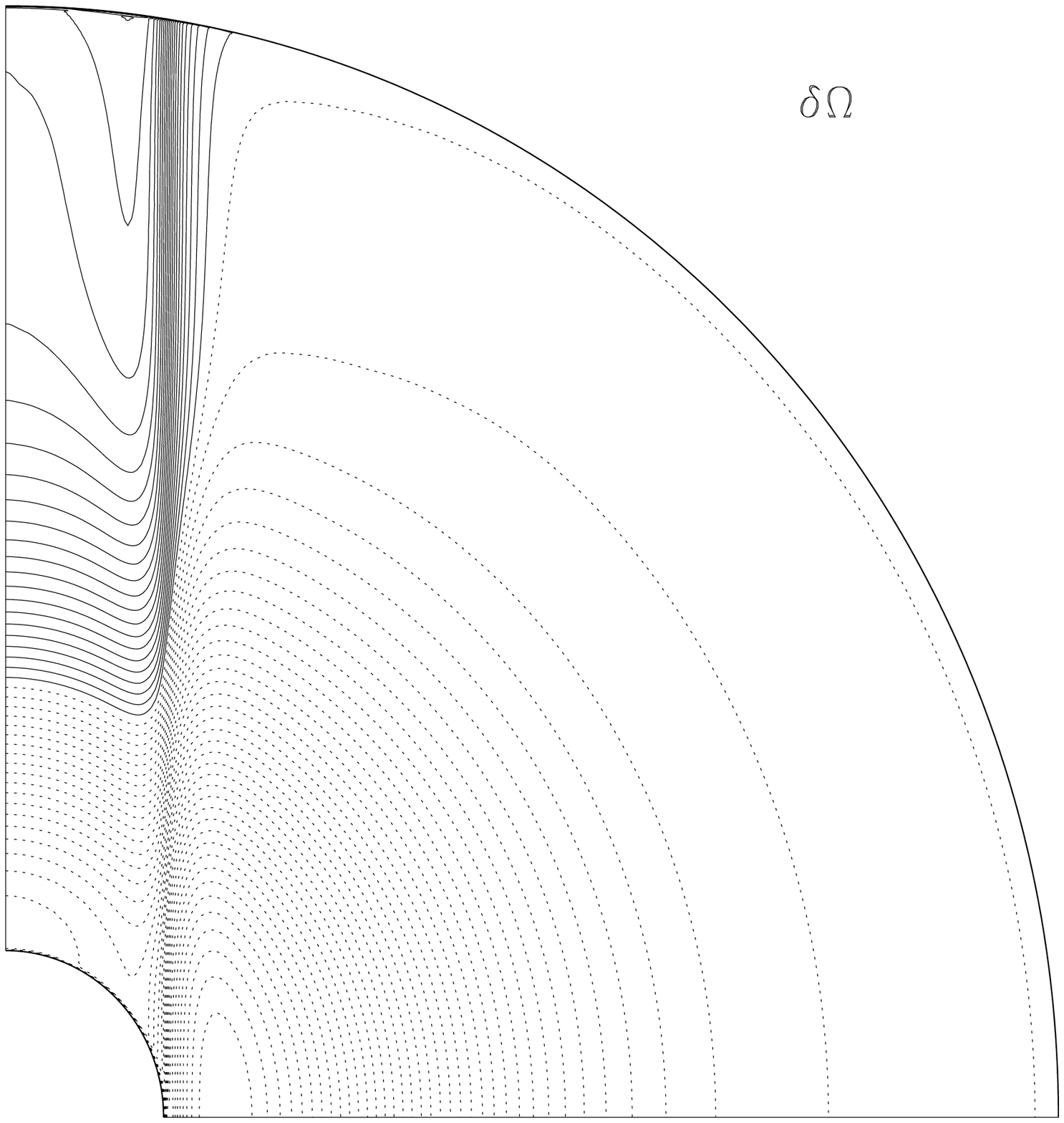}
 \includegraphics[width=0.3\textwidth,height=0.216\textheight,angle=0]{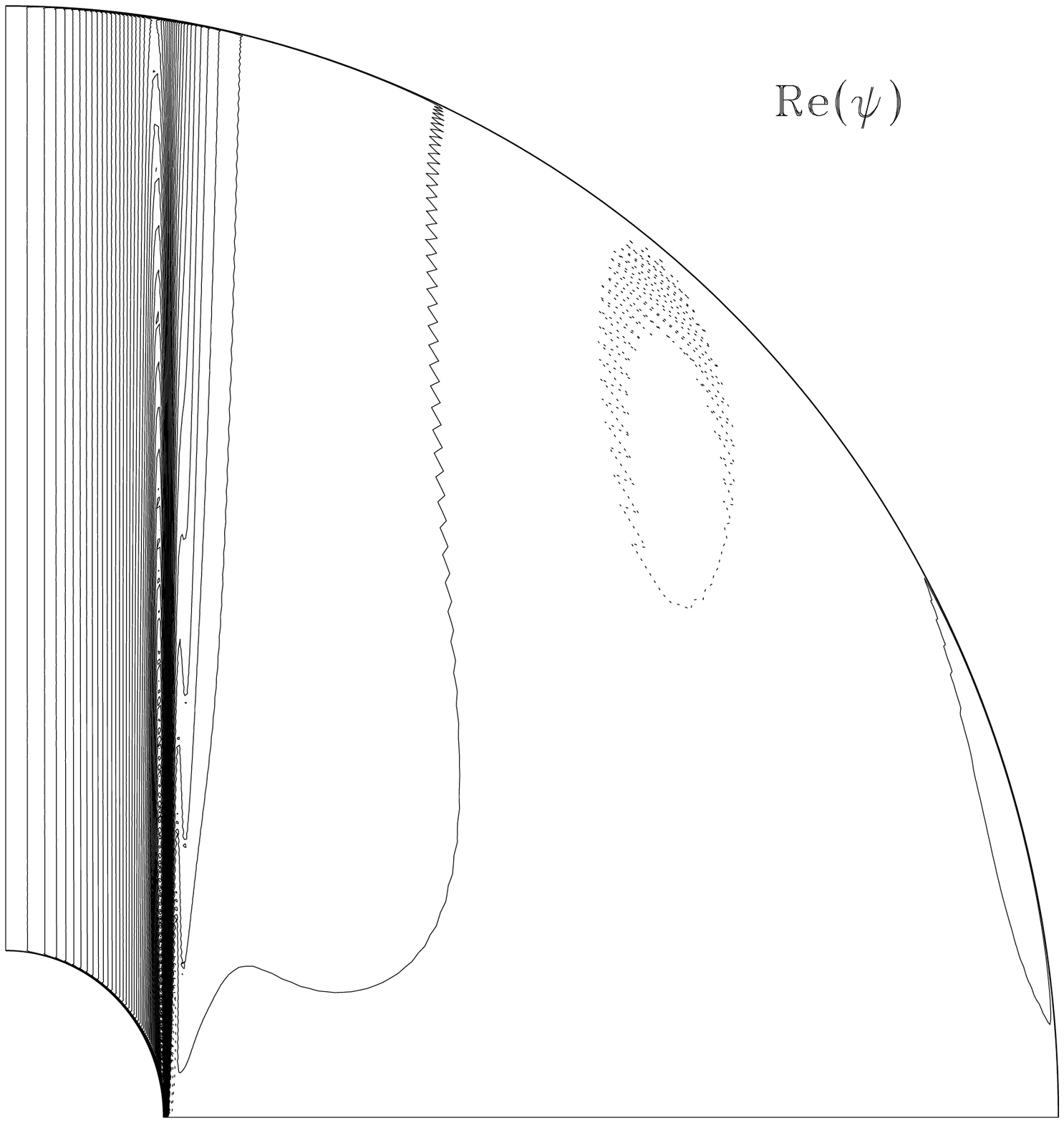}\\
 \end{center}
\caption{Evolution of the differential rotation $\delta \Omega$,
 the meridional circulation $\psi$ for $\eta=0.15$, $\tilde\rho=10$,
 $E=10^{-7}$, $\lambda=10^{-4}$ and $B=0,10^4,10^5,10^7$ (downward).
 In the first column, difference in differential rotation is shown with
 contours : solid ones are faster than the
 core and dashed contours are slower than the core.
 When the spin-up flow dominates the differential rotation 
 we get a fast equator while baroclinicity induces a fast pole.
 In the second column, meridional circulation is described with dotted lines for
 clockwise circulation (solid lines for counter-clockwise circulation).  
 Boundary conditions are no-slip on both
 sides $r=\eta$ and $r=1$. \label{fig5}}
\end{figure*}

When gravitational contraction occurs in a baroclinic envelope, two
drivings compete : the baroclinic torque and the one induced by mass
contraction. To get a global view of this competition we therefore need
resorting to numerical solutions. The numerical method is detailed in
appendix.  The global problem is the superposition of the two flows :
the one induced by mass contraction (\ref{eq12}) and the baroclinic one
(\ref{eq9}).  Since the system is linear, the full solution is a linear
combination of both. At the equator it reads

\begin{eqnarray}
u_\phi(r,\theta=\frac{\pi}{2})=\sqrt{\frac{2}{E}} (1-r^2)^{3/4}\left(
\frac{\eta^2}{r} - \frac{9}{4\tilde \rho \eta} r\right) \nonumber \\
-r B \int_{r}^{1} \frac{n^2(r)}{r} dr \; .
\label{eq16}
\end{eqnarray}
From the foregoing equation, we see that the differential rotation is
governed by baroclinicity when

\begin{equation}
 B \gg E^{-1/2}\; .
 \label{eq14}
\end{equation}
It is shown in Fig.~\ref{fig1}.
However, we also know (from the angular
momentum flux balance of a steady flow, see \citealt{R06}) that the
meridional circulation associated with baroclinic flows is \od{BE},
while the meridional circulation of the contraction-induced spin-up
is \od{1}. Hence a baroclinic flow completely controls the dynamics
as long as $BE\gg1$. Note that this inequality implies \eq{eq14} since
$E\ll1$. When the spin-up strengthens, the foregoing inequalities predict
an intermediate regime

\begin{equation}
E^{-1/2} \ll B \ll E^{-1}
\end{equation}
where the differential rotation is of baroclinic origin ($BE^{1/2}$) but
the meridional circulation is driven by contraction ($BE\ll1$). Finally,
when $BE^{1/2}\ll1$, the flow is fully controlled by the contraction
induced spin-up.

\begin{figure}
 \begin{center}
 \includegraphics[width=0.3\textwidth,height=0.135\textheight,angle=0]{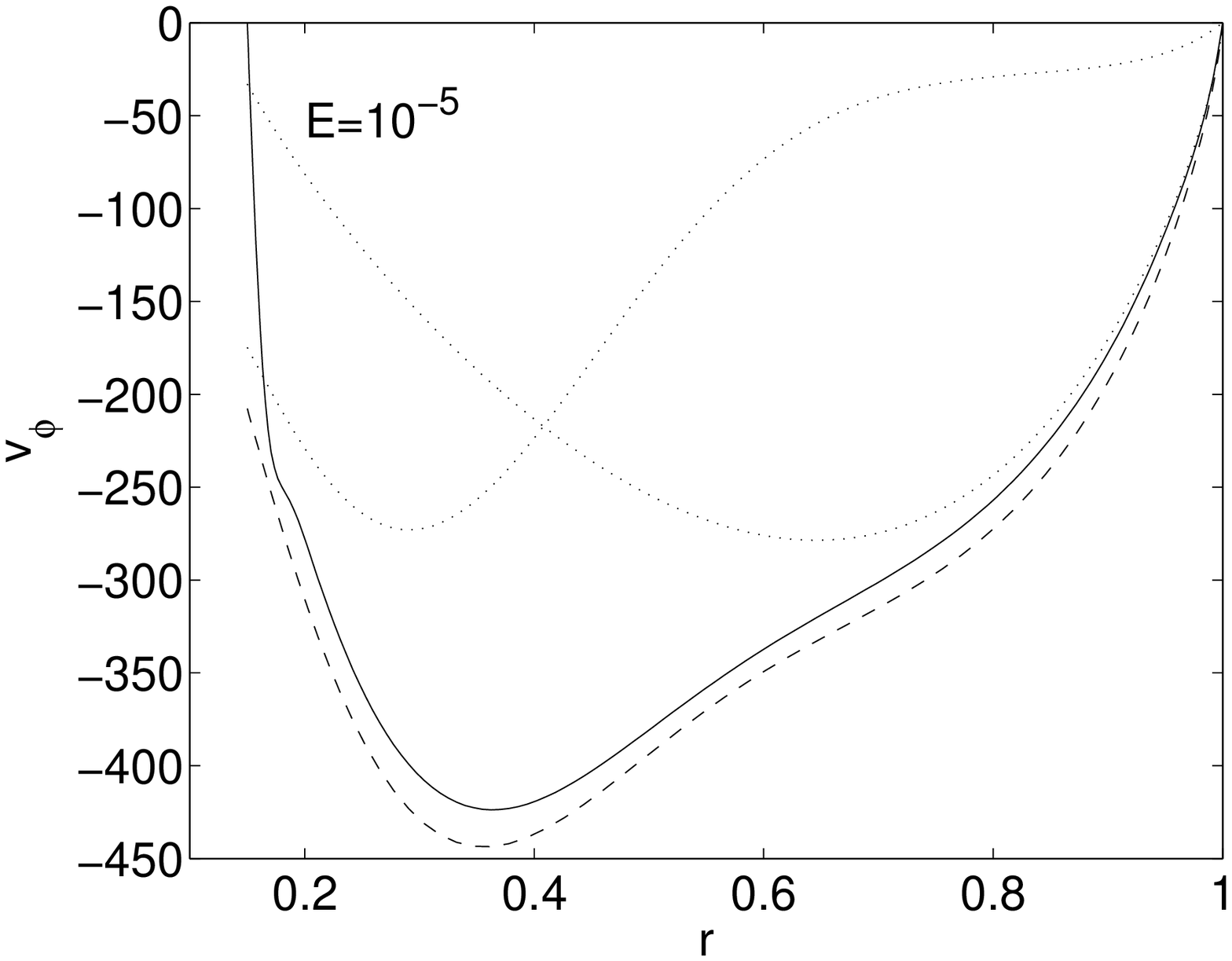}
 \includegraphics[width=0.3\textwidth,height=0.135\textheight,angle=0]{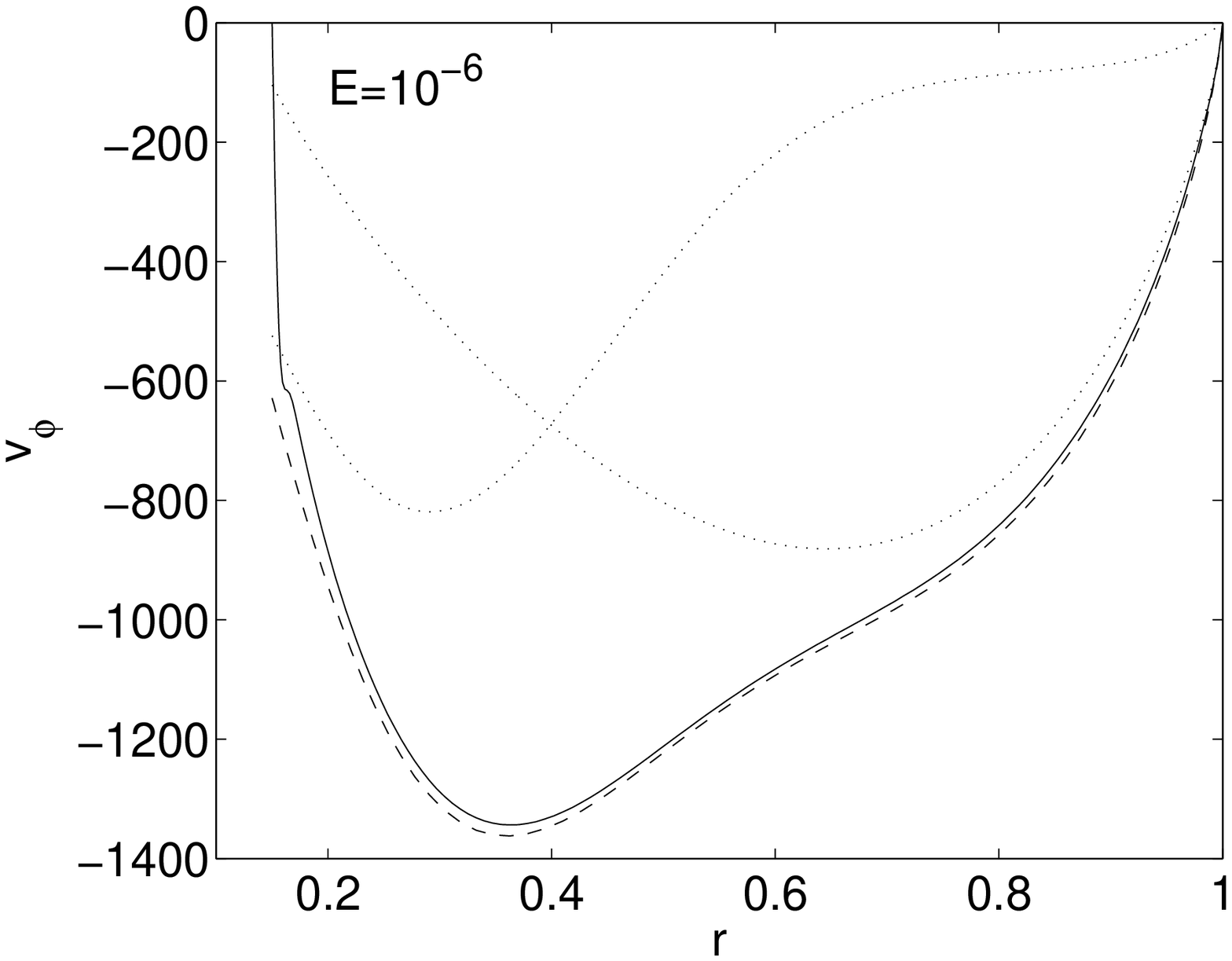}
 \includegraphics[width=0.3\textwidth,height=0.135\textheight,angle=0]{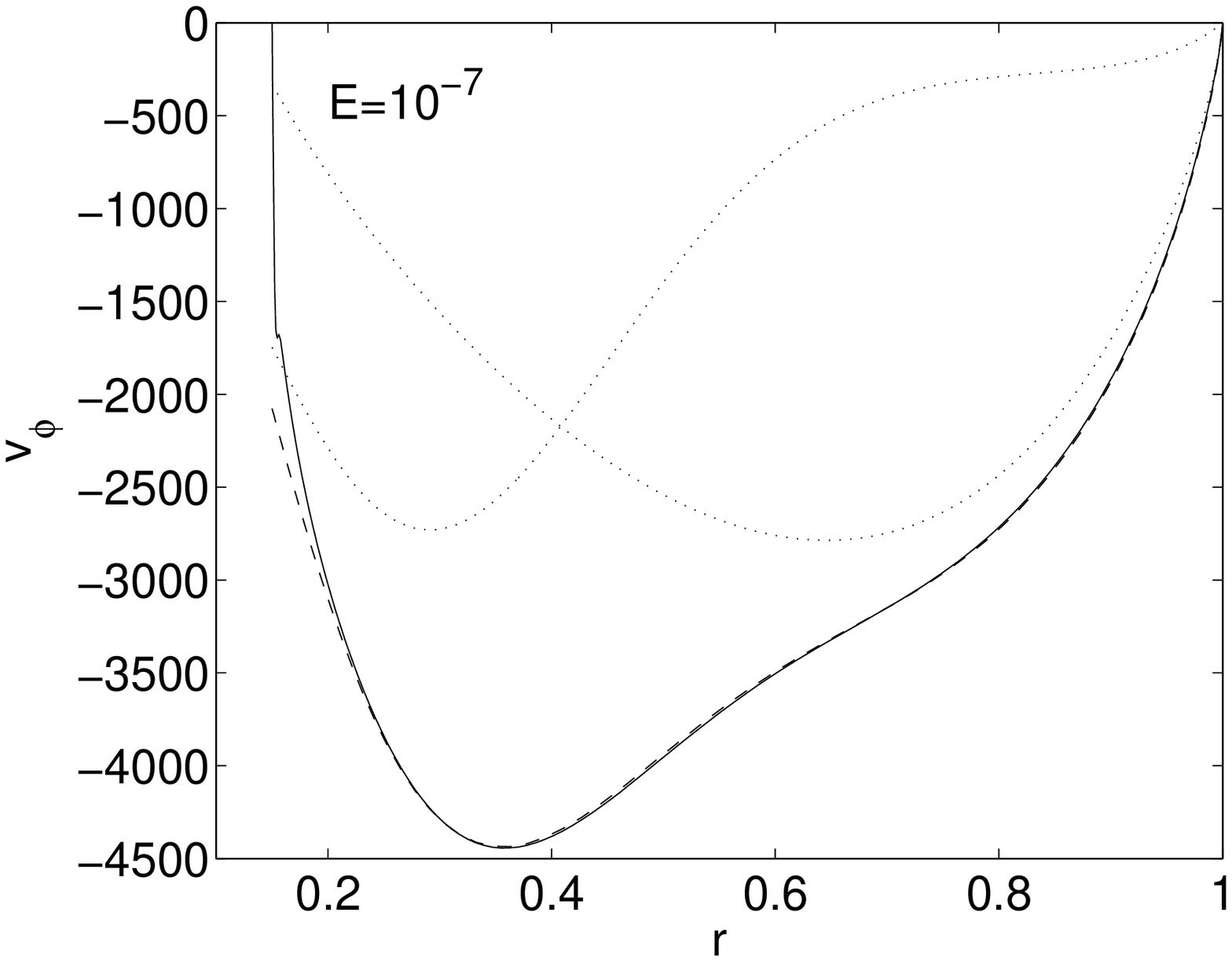}
\end{center}
\caption{$u_\phi$ as a function of the normalized radius for $\lambda=10^{-4}$,
$\eta=0.15$ and $\tilde\rho=10$.  The solid line is the numerical
solution, dotted lines are the analytical solutions of the spin-up
flow and the baroclinic one, the dashed line is the sum of both.
Downward : $E=10^{-5},10^{-6},10^{-7}$ and the transition value on
the differential rotation is respectively $B=10^3,3.10^3,10^4$.
Beyond this treshold, the baroclinicity governs the differential rotation.
Boundary conditions are no-slip on both sides $r=\eta$ and $r=1$.
Note that as the Ekman number is getting smaller,
the discrepancy between the numerical and the analytical solutions
does too.  \label{fig6}}
\end{figure}

This two step transition is confirmed by numerical solutions as
illustrated in Fig.~\ref{fig5} and Fig.~\ref{fig6}. There, for a given
$E$, $B$ is progressively increased and we clearly see the intermediate
regime where differential rotation and meridional circulation are of
different origin (third row).

\section{The case of stress-free boundary conditions}

The outer layers of the envelope are not rigidly attached to the core.
Therefore the use of outer stress-free boundary conditions is
more realistic.  In this case however, we no longer have access to an
analytical expression of the flow in the envelope and have to resort to
numerical solutions.

\subsection{Scaling the steady mass contraction induced flow}

\begin{figure}
 \begin{center}
 \includegraphics[width=0.3\textwidth,height=0.216\textheight,angle=0]{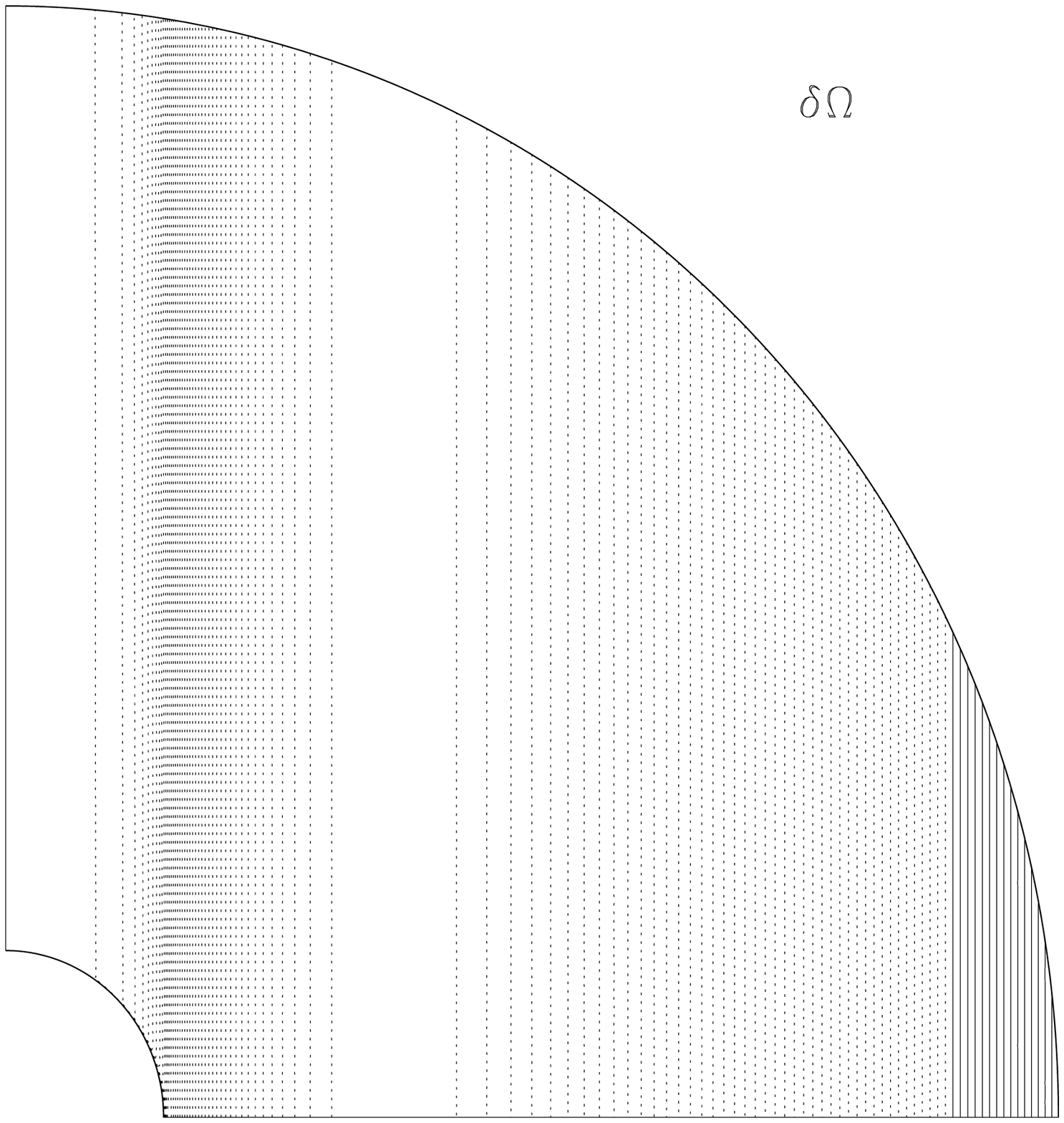} 
 \includegraphics[width=0.3\textwidth,height=0.216\textheight,angle=0]{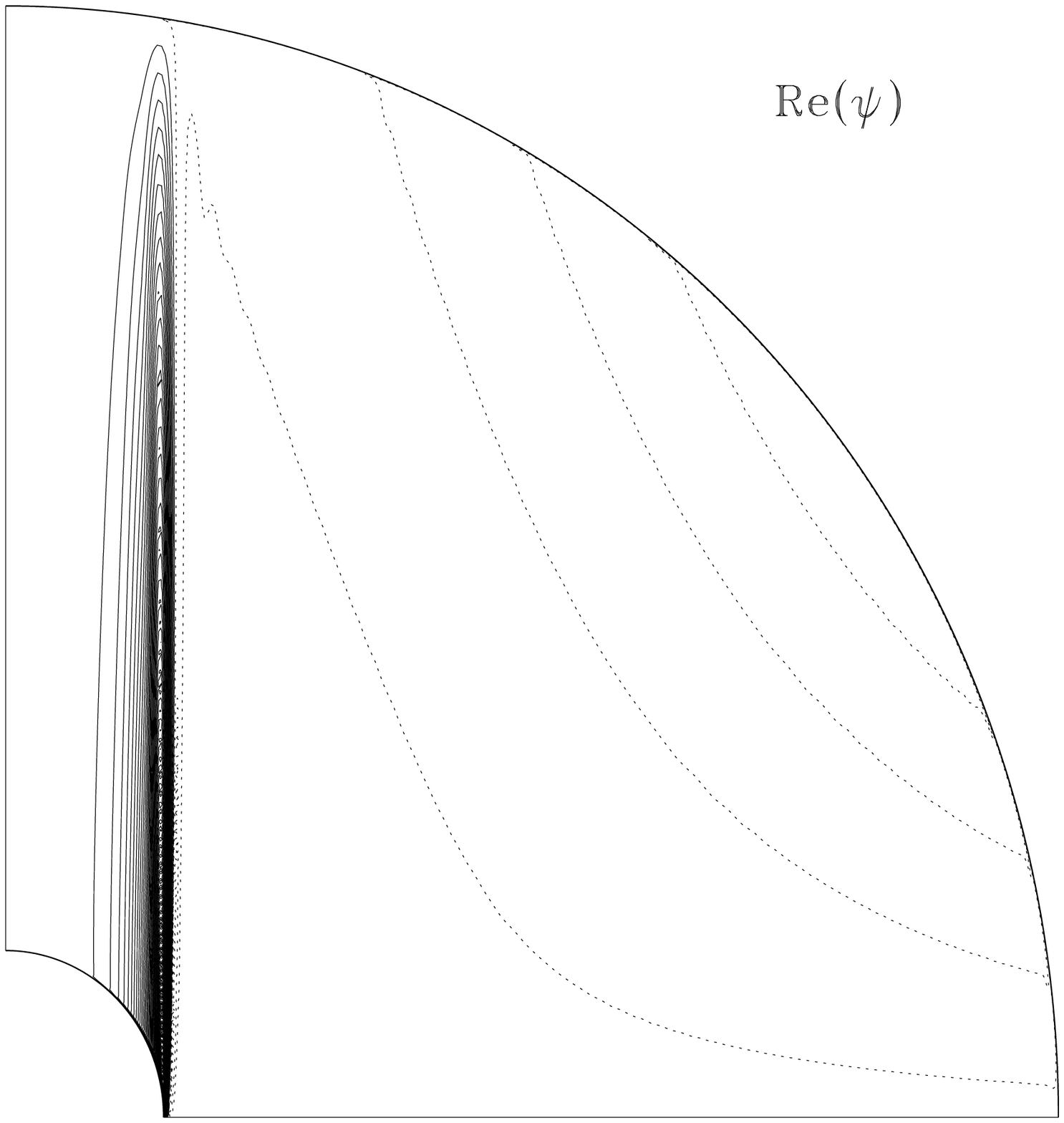}\\
 \end{center}
\caption{Differential rotation and meridional circulation for $E=10^{-7}$,
$\eta=0.15$ and $\tilde\rho=10$ for an unstratified configuration.
For $\delta \Omega$, dashed (resp. solid) lines represent
rotation slower (resp. faster) than the core. For $\psi$, dotted (resp. solid) lines are for
 clockwise (resp. counter-clockwise) circulation.
Boundary conditions are no-slip at $r=\eta$ and stress-free
at $r=1$.\label{mercut}}
\end{figure}

Let us first study the steady solution of the spin-up flow.  As shown
in Fig.~\ref{mercut}, it exhibits the typical
cylindrical differential rotation of a dominating mass contraction flow.
The equatorial surface region rotates faster than the core and the
pole is slower.  The meridional circulation displays two cells with a
strong Stewartson layer at $s=\eta$ (compared with the previous with no-slip
conditions).

\begin{figure}
 \begin{center}
 \includegraphics[width=0.5\textwidth,height=0.25\textheight,angle=0]{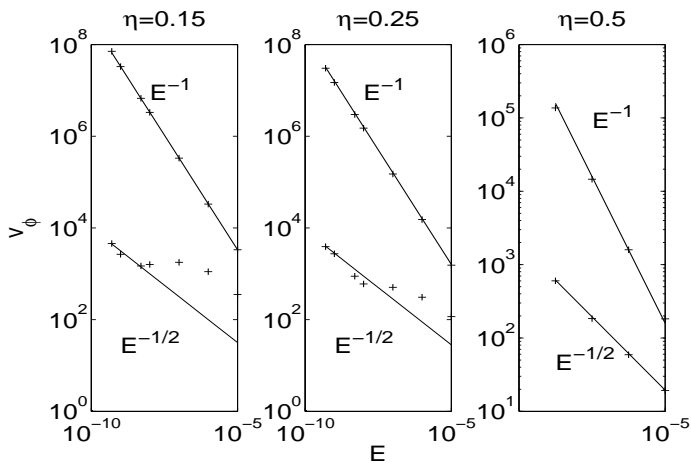}\\ 
 \end{center}
\caption{Logarithm of the absolute value of the amplitude of the numerical
azimuthal velocity as a function of the logarithm of the Ekman number
at $z=0.5$.  The left pannel is for $\eta=0.15$, the middle one is
for $\eta=0.25$ and the right one for $\eta=0.5$.  Inside the tangent
cylinder, measures were taken respectively at the points $s=0.1,0.2,0.25$,
the azimuthal velocity scales around $E^{-1/2}$.  It is proportional
to $E^{-1}$  within the envelope $s=0.65$.  Note that the azimuthal
velocity is mainly negative for small cores and positive for bigger
cores like $\eta=0.5$.
\label{fig10}}
\end{figure}

In Fig.~\ref{fig10}, we show the amplitudes of the mass contraction
flow at two positions : inside and outside of the tangent cylinder.
Since the boundary conditions on the core are no-slip, the flow within
the tangent cylinder is expected to be $\mathcal{O}(E^{-1/2})$ according
to expression (\ref{Usol}).  For small cores $\eta=0.15$ or $0.25$, numerical
solutions show that
the Stewartson layer impacts the interior of the tangent cylinder and
that the asymptotic state is reached only at extremely small values of the
Ekman number $E \leq 10^{-8}$. When the core is bigger, for
instance $\eta=0.5$, the $E^{-1/2}$ scaling inside the tangent cylinder
is clearly showing up for all Ekman numbers less than $10^{-5}$.

Outside the tangent cylinder, numerics show that the differential rotation
is always $\mathcal{O}(E^{-1})$ when the outer boundary conditions are
stress-free.  Such an important amplitude indicates that the steady state
may not be reached during the contraction phase and may not be studied
with linear equations since quadratic terms are expected to be important,
namely $\mathcal{O}(\frac{Ro}{E^2})$.

\subsection{The transient phase}

The large amplitude of the steady state outside the tangent cylinder
forces us to consider the time evolution of the solution of the mass
contraction induced flow.  To do so, we solve the set of equations
(\ref{eq4}) with an order one implicit scheme (Euler's method) so as
to eliminate inertial waves and concentrate on the secular evolution.

\begin{figure}
 \begin{center}
 \includegraphics[width=0.5\textwidth,height=0.25\textheight,angle=0]{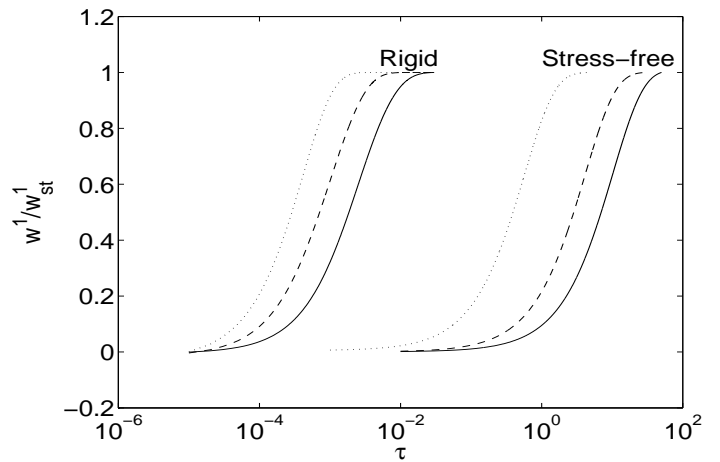}\\ 
 \end{center}
\caption{Time evolution of the major component of the azimuthal velocity
$w^{l=1}$ normalized by its steady solution $w^{l=1}_{\text{st}}$ as
a function of non-dimensional time with $E=10^{-5},10^{-6},10^{-7}$
(respectively the dot, the dashed and the solid line) and $Ro=10^{-6}$
at the radius $r=0.88$.  On the left side are the solution with rigid
boundary conditions on both sides and on the right side the solutions
with stress-free boundary conditions on the outer.  In the first case,
the steady state is reached in a time smaller the dimensionless time
of contraction, while in the second case this time is longer and scale
as $E^{-0.86}$.
\label{fig11}}
\end{figure}

In Fig.~\ref{fig11}, we plot a proxy of the amplitude of the differential
rotation for various Ekman numbers and for no-slip and stress-free outer
boundary conditions. With no-slip conditions, we see that the steady
spin-up flow is quickly established and justifies the use of a steady
solution. On the contrary, the use of outer stress-free boundary
conditions leads to a much longer transient flow that lasts more than the
typical time scale of the driving by gravitational contraction.

To go further it is interesting to characterize this transient flow with
respect to the parameters of the problem. From the numerical solution we
find that the transient duration $\tau_{\rm sf}$ scales like

\beq
 \tau_{\text{sf}}\propto Ro E^{-0.86} 
\eeqn{tauf}
This scaling of the Ekman number is very close to $E^{-6/7}$ that is
reminiscent of one of the scalings of the Stewarston layer in the
spherical Couette flow \cite[see][]{stewar66}. In these layers that
surrounds the core along the tangent cylinder, a typical thickness is
$E^{2/7}$. This might control the amplitude of the flow outside the
tangent cylinder when stress-free outer conditions are met. The analysis
of the Stewartson layer associated with this transient flow
is difficult and beyond the scope of the present work.

\begin{figure*}
 \begin{center}
 \includegraphics[width=0.26\textwidth,height=0.216\textheight,angle=0]{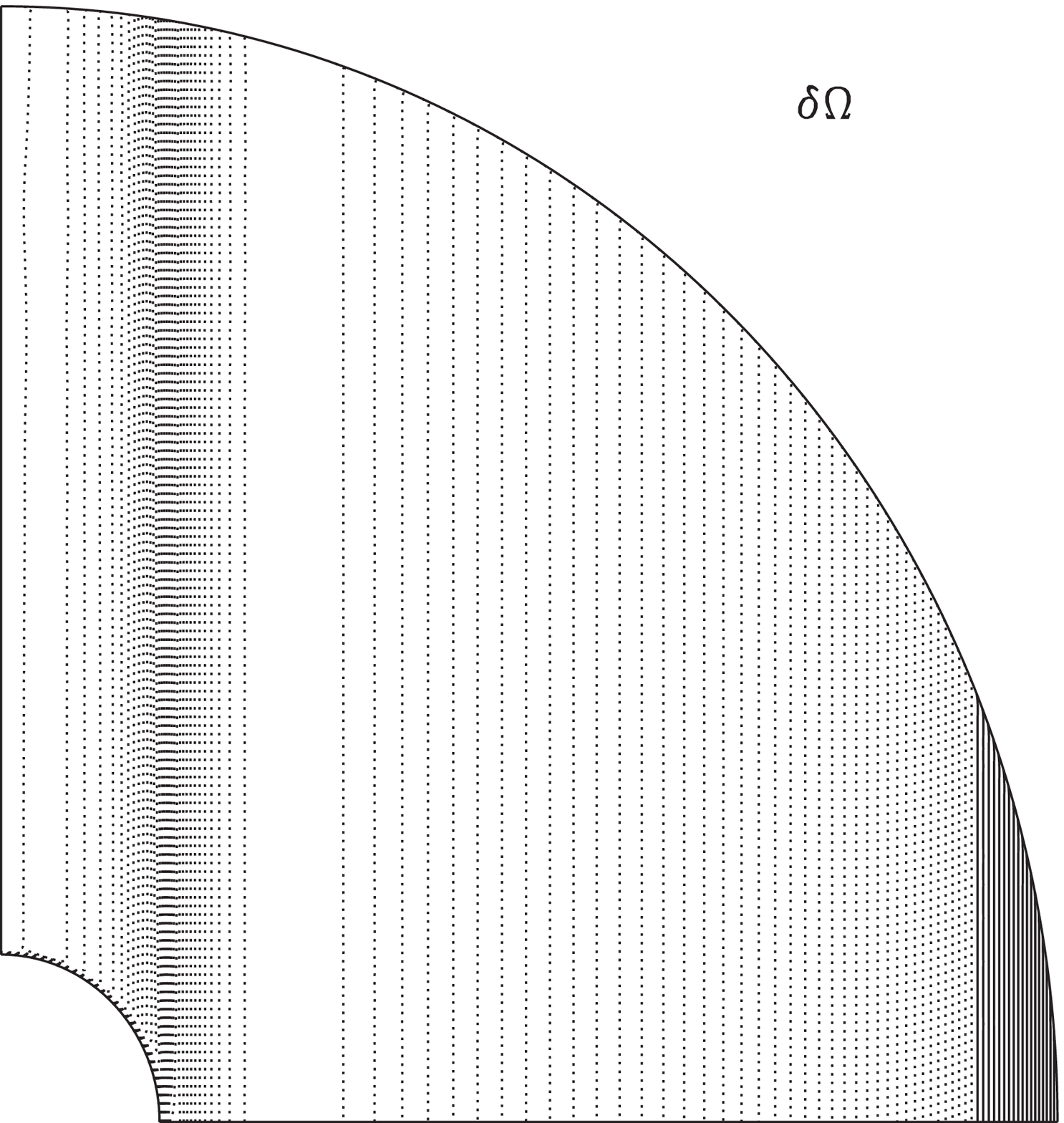} 
 \includegraphics[width=0.48\textwidth,height=0.216\textheight,angle=0]{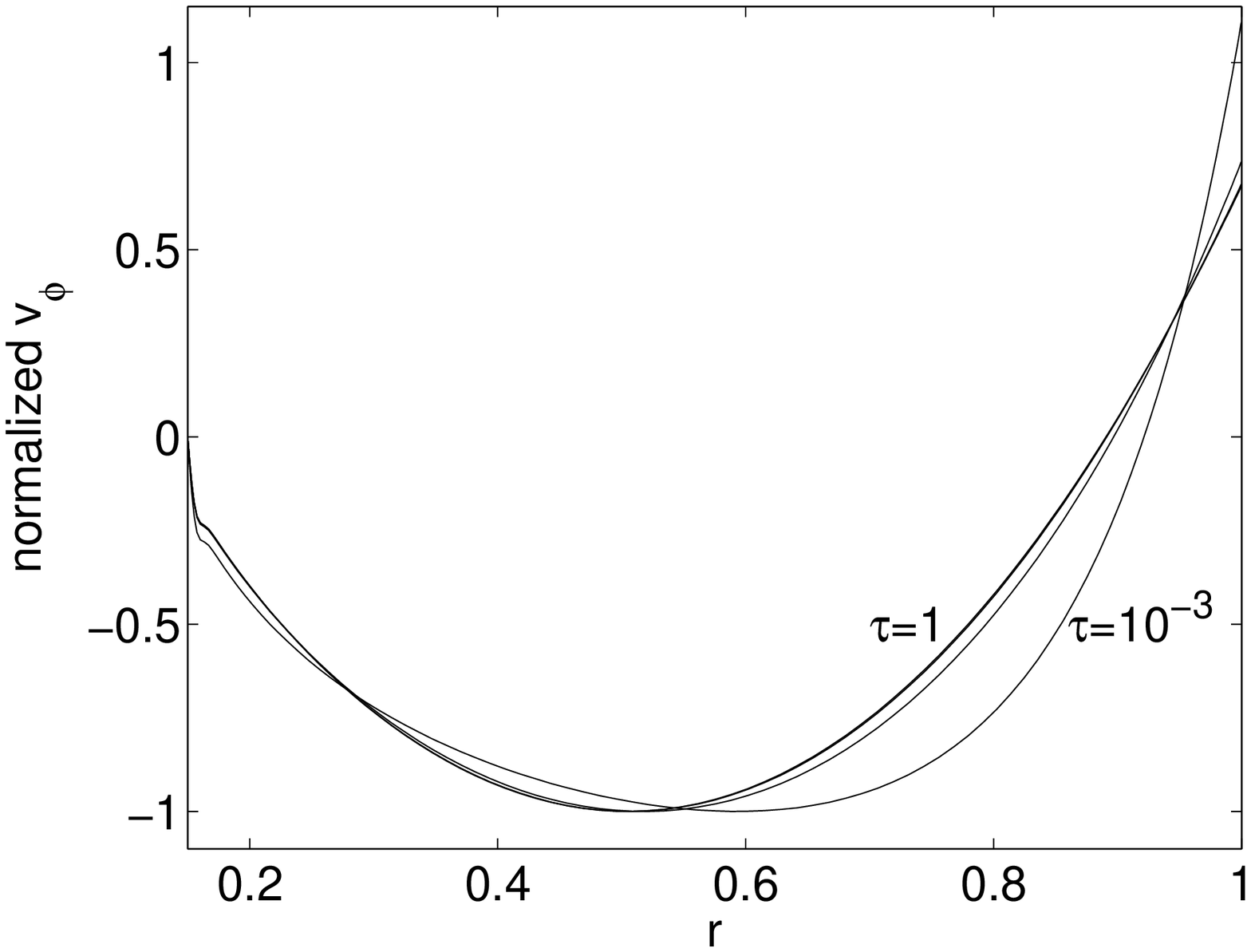}\\
 \end{center}
\caption{Left: the differential rotation associated with the transient
phase of a spin-up induced by mass contraction for $E=10^{-6}$ and
$Ro=10^{-6}$ at time $\tau=0.02$. Dashed (resp. solid) lines represent
rotation slower (resp. faster) than the core. Right: the profile of the
corresponding azimuthal velocity in the equatorial plane as a function
of the normalized radius for various times. The curves are scaled by
the absolute value of the amplitude at the minimum.
\label{figprofil}}
\end{figure*}

Another remarkable property of the transient flow is its approximate
self-similarity. Its spatial shape remains almost unchanged, while its
amplitude grows as time passes. The associated differential rotation is
parallel to the $z$-axis as shown in Fig.\ref{figprofil}.  Its amplitude
grows according to the time profile displayed in Fig.\ref{fig11}.
This time dependence can be approximated as

\beq \frac{A}{E}(1-e^{-\tau \ln10/\tau_{\text{sf}}})\eeqn{lf}
where $A$ is a constant of order unity.

The foregoing result may be translated in the stellar case. It shows
that a contracting, fully convective star, may reach a self-similar
spin-up flow with cylindrical rotation. Neglecting viscous force (in fact
Reynolds stresses), we may expect that substituting $\rho\vv$ to the
foregoing incompressible velocity field, we can get an good representation
of the actual flow in a compressible envelope. This is supported by the
fact that the geostrophic balance is unchange in this case. Of course,
this conjecture has to be verified by the study of the compressible case.

\subsection{Transient phase and stratification}

During the contraction of a star on the PMS, an initial convective
envelope progressively leaves the place to a radiative envelope when the
star is massive enough. This radiative envelope is stably stratified and
without any extra-forcing from gravitational contraction it would relax
to the steady baroclinic state that we mentioned before.

One question is therefore whether the contraction induced spin-up is
strong enough to overwhelm the foregoing baroclinic flows that are
themselves transient flows. To have an idea of the result we may use the
amplitude of a steady baroclinic flows as an upper limit of the actual
flows. In this perspective we can use the results of \cite{ELR07} and
\cite{ELR13}
who showed that the baroclinic flow
in a radiative envelope is characterized by a differential rotation that
is typically 15\% of the bulk rotation. Let us assume that the maximum
amplitude of the baroclinic flow reads

\[ V_b = k V_{\rm eq}\]
where $k\infapp 0.2$. During the phase of gravitational contraction
the spin-up flow grows according to \eq{lf}. At some time $\tau_c$,
the spin-up flow overwhelms the baroclinic flow. At such time we have

\[ \frac{A}{E}(1-e^{-\tau_c \ln10/\tau_{\text{sf}}}) = \frac{V_b}{V_{s}}
\]
with $V_{s}\sim R/t_{\rm KH}$, $t_{\rm KH}$ being the
Kelvin-Helmholtz time of the star. Hence,

\beq \frac{\tau_c}{\tau_{\text{sf}}} \sim -\log\lp1 - \frac{EV_bt_{\rm
KH}}{AR}\rp\eeqn{tauc}
Using numbers of a typical 3~M$_\odot$ star on the birthline, we find that

\[ \frac{EV_bt_{\rm KH}}{AR} = \frac{k}{2A}\frac{\nu GM^2}{R^3L}\infapp
10^{-4}\]
This small ratio indicates that $\tau_c\ll\tau_{\text{sf}}$ so that the
contraction-induced flow overwhelms the baroclinic one during the linear
growth of the transient flow \eq{lf}. Hence, from \eq{tauf} we get

\[ \tau_c \sim \frac{E^{0.14}Ro}{A}\frac{V_bt_{\rm KH}}{R} \; .\]
With the definition of $Ro$ and $V_b$ it turns out that 

\[ \tau_c \sim \frac{k}{2A} E^{0.14} \; .\]
Because of the very small value of the Ekman number, $\tau_c$ is clearly
less than unity showing that the spin-up flow will in the end take over
the baroclinic flow, likely much before this latter flow can be established.

\begin{figure}
 \begin{center}
 \includegraphics[width=0.45\textwidth,height=0.25\textheight,angle=0]{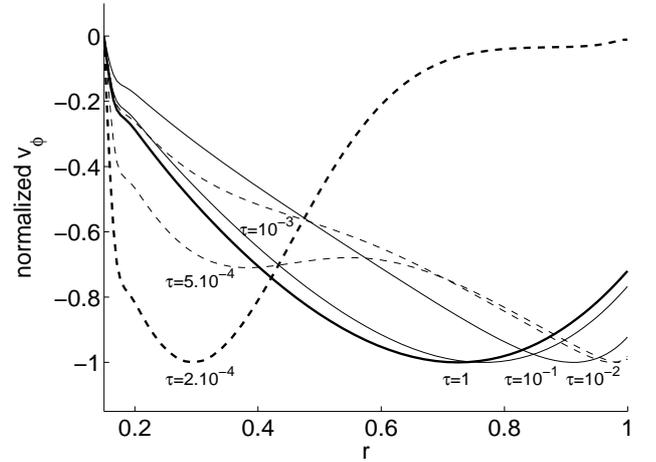}\\
 \end{center}
\caption{Radial profiles taken at various times of the azimuthal
velocity in the equatorial plane of the star for $E=10^{-5}$, $B=10^2$,
$\epsilon=10^{-7}$, $\lambda=10^{-4}$ and $Ro=10^{-5}$.  Profiles are
scaled as in Fig.~\ref{figprofil}. The dimensionless time $\tau$ is
specified for each curve.  Dashed lines are profiles for baroclinicly
dominated dynamics while solid ones are for spin-up dominated dynamics.
The dashed bold curve shows the initial profile and the solid bold curve
shows the last one.}
\label{figspinstrat}
\end{figure}

We verified this conclusion with a numerical simulation integrating the
spin-up flow from a pre-existing baroclinic flow driven by a fixed stable
stratification.  In Fig.~\ref{figspinstrat}, we show the time evolution of
the azimuthal velocity in the equatorial plane of the star.  It describes
the transient phase from a steady baroclinic flow to a growing spin-up
flow.  The transition between the two flows happens between $\tau=10^{-3}$
and $\tau=10^{-2}$. This is less than $\tau_c\sim 0.2$ (at $E=10^{-5}$),
namely less than our first evaluation obtained by a comparison of the
amplitudes of the flows (see Eq.~\ref{tauc}). The parameters have been
chosen such that $BE\ll1$ as expected in real situations. $\tau_c$
therefore appears as a good indicator of the time needed
by spin-up to overwhelm baroclinic flows.  Let us finally note that, once
the spin-up flow is settled, the flow remains approximately self-similar
and this for at least 80\% of the Kelvin-Helmholtz time.

\section{Discussion and conclusions}

The gravitational contraction that occurs before or after the main
sequence is strongly influencing the rotation
rate of the stars and their internal dynamics. In the foregoing study,
we have investigated the consequences of the combination of rotation
and gravitational contraction with a very simplified model in order to
decipher this complicated dynamics and be prepared for the construction
of more elaborated models of rotating stars like the ESTER models of
\cite{ELR13}.

Thus, to the compressible gas of the star, we substituted an
incompressible fluid that may also be stably stratified. Schematically,
the bell-shaped profile of the star density is replaced by a step
function delineating a central core that absorbs matter from the
envelope as the star contracts. The size of the core is small and
arbitrary. The envelope is either neutrally or stably stratified.

During the contraction on the PMS the stellar envelope of an
intermediate-mass star usually passes from a convective to a radiative
state. But the radiative state is stably stratified. The combined effect
of rotation and stable stratification drives a baroclinic flow that may
contest a pre-existing spin-up flow built up during a previous convective
phase of the envelope. The question is which of these flows will govern
the dynamics of the contracting star and finally determine the initial
conditions of the dynamics on the main sequence.

Using our simplified model we compared the strength of these two flows
and found that although contraction is slow, the induced spin-up controls
the large-scale flows of the outer envelope, namely its differential
rotation and meridional circulation. Moreover, our model underlines
the role of the outer boundary conditions and shows that with realistic
stress-free conditions we should expect an unsteady flow.  In addition,
it shows that this transient flow keeps a self-similar shape during its
growth (if we omit boundary layers).

When the star reaches the main sequence, the contraction turns
off and the flows in the envelope relax towards the steady baroclinic
flow on the Eddington-Sweet time scale. As far as intermediate-mass stars
are concerned, because of their fast rotation, the Eddington-Sweet time
scale is close to the Kelvin-Helmholtz one and the transition to the
quasi-steady state of the main sequence is quite short (for instance for
7~\msun\ star, the Eddington-Sweet timescale is 2.8~Myr for a rotation near
break-up, to be compared to the 46~Myr of the lifetime of such stars).
On the other hand, if for some reason (mainly the combination of magnetic
fields and mass loss), the star loses much angular momentum so that
$2\Omega\ll N$ at the beginning of the main sequence, then the dynamic
state on an outer envelope will be controlled by slowly decaying
baroclinic modes excited by the contraction phase. The slow decay may
occupy a significant fraction of the main sequence with
consequences on the mixing processes.

Back to fluid dynamics, the simple model that we used shows other details
on the dynamics of this system, like for instance the shear layer (the
Stewartson layer) that circumvent the core on its tangent cylinder.
Such a feature is clearly an artefact of the model for stars with no
convective cores like PMS stars, but it is certainly an important feature
for stars leaving the main sequence where the core contracts and the
envelope expands. At the core-envelope interface the build-up of a
density jump due to nuclear evolution, combined with rotation, triggers
a Stewartson layer on the tangent cylinder of the discontinuity.
This layer may indeed explain the efficient transport
of angular momentum between the core and envelope of giant or subgiant
stars that is needed to explain the rather mild radial differential
rotation observed in these stars
\cite[][]{deheuvels_etal12,mosser_etal12,eggenberger_etal12}.
Indeed, let us remark that our model shows that there is a tight coupling
between the inner part of the tangent cylinder and the core itself. This
coupling is essentially a consequence of the Taylor-Proudman theorem that
imposes no velocity gradient along the rotation axis (columnar flow).
Therefore, the transport of angular momentum between the core and
the envelope is much enhanced by the Stewartson layer. Since such a
layer has a surface that is $R_{\rm star}/R_{\rm core}$ larger than
the surface of the core, we expect that the flux of angular momentum
between the core and the envelope will be enhanced by a similar factor
(viscosity and velocity gradient being assumed similar) with respect to
a 1D shellular profile. Noting that the moment of inertia of the core
and of the matter inside its tangent cylinder are not much different,
we expect that the Stewartson layer plays a crucial role in the angular
momentum exchange between core and envelope and might well be the key
feature that reconcile models and observations. It is clear that present
1D models do not take this fluid dynamics feature into account and that
the final answer will be given by 2D models incorporating this flow. A
dedicated study is clearly needed to give a quantitative estimate of
this effect and to offer a new comparison with observations.

Hence, more than the numbers and the applicability to a given object,
the foregoing Boussinesq model underlines the main features of the
dynamics of a contracting and rotating envelope. It stresses the
key role of outer boundary conditions and the various flows that might
govern a contracting phase depending on the strength of the
stratification. The side effect of the core in this model underlines the
role of a Stewartson layer that may appear either after a rapid
change in density or in viscosity. The model also stresses the fact that
no steady state can be expected as for the interior flows but that these
flows may converge towards a universal one when gravitational contraction
ceases. The next step of these investigations focusing specifically
on stars will be developed with the full physics using the ESTER code
of \cite{ELR13}.

\begin{acknowledgements}
We would like to thank the referee for his constructive remarks
on the first version of the manuscript. We also acknowledge the support
of the French Agence Nationale de la Recherche (ANR), under grant ESTER
(ANR-09-BLAN-0140).  The numerical calculations have been carried out
on the CalMip machine of the `Centre Interuniversitaire de Calcul de
Toulouse' (CICT) which is gratefully acknowledged.
\end{acknowledgements}

\bibliographystyle{aa}
\bibliography{bibnew}

\appendix
\section{Numerical method}

To solve the set of linear equations (\ref{eq5}), we discretize the equations
using a spectral
method. We project the velocity field onto the harmonics spherical base 
\begin{equation}
\vec{u}=\sum \limits_{l=0}^{+\infty} {\sum \limits_{m=-l}^{+l} {u_m^l
\vec{R}_l^m + v_m^l \vec{S}_l^m + w_m^l \vec{T}_l^m }}
\end{equation}
where
\begin{equation}
\vec{R}_l^m=Y_l^m \vec{e_r},\qquad \vec{S}_l^m=\vec{\nabla}Y_l^m, \qquad
\vec{T}_l^m=\vec{\nabla}\times \vec{R}_l^m
\end{equation}
$Y_l^m$ are the normalized spherical harmonics, $\vec{e_r}$ is the
radial unity vector and $\vec{\nabla}$ is defined on the unity sphere.
We write the temperature perturbation onto the spherical harmonics base too :

\begin{equation}
 \theta_T = \sum \limits_{l=0}^{+\infty} {\sum \limits_{m=-l}^{+l} t_m^l
Y _l^m}
\end{equation}
We add finally the boundary conditions on this field :

\begin{equation}
\left \{
\begin{array}{ccl}
t_l'(r=\eta) = 0\\
t_l(r=1)=0\\
\end{array}
\right .
\end{equation}

We discretize the radial direction for $r \in [\eta;1]$ onto the
Gauss-Lobatto grid associated with the Chebyshev polynomials.
Thereby, equations are solved in two dimensions $(r,\theta)$.
The system is axisymmetric which implies $m=0$.\\
The equation of continuity reads

\begin{equation}
v_m^l= \frac{1}{\Lambda } \frac{1}{r}\frac{\partial}{\partial r}(r^2
u_m^l)
\label{eqcont}
\end{equation}
where $\Lambda=l(l+1)$.\\
The energy equation reads

\begin{equation}
 B \tilde E_T (r^2 \frac{\partial^2}{\partial r^2}t^l_m +2r \frac{\partial}{\partial r}t^l_m  -\Lambda t^l_m) = n_T^2(r) r^2 u^l_m
\end{equation}
The equation of motion is projected onto two directions because
equations on $\vec{R}_l^m$ and on $\vec{S}_l^m$ are redundant.\\
On $\vec{R}_l^m$, it reads

\begin{eqnarray}
A_{l-1}^l r^{l-1}\frac{\partial}{\partial r} \left ( \frac{u_{m}^{l-1}}{r^{l-2}} \right ) +A_{l+1}^l r^{-l-2}\frac{\partial}{\partial r}\left ( r^{l+3}u_{m}^{l+1} \right)\nonumber \\
+E \Delta_{l} w_{m}^{l}=-\sqrt{\frac{4\pi}{3}}\delta_{l1} (\frac{\eta^2}{r^2} - r\dot\omega)
\end{eqnarray}
$\delta_{ij}$ is the Kronecker symbol.\\
On $\vec{T}_l^m$, it reads

\begin{eqnarray}
-B_{l-1}^l r^{l-1}\frac{\partial}{\partial r} \left ( \frac{w_{m}^{l}}{r^{l-1}} \right ) -B_{l+1}^l r^{-l-2}\frac{\partial}{\partial r} \left ( r^{l+2}w_{m}^{l+1} \right )\nonumber \\
-l(l+1)t_{m}^{l}+ E \Delta_{l} \Delta_{l}(ru_{m}^{l})=-\sqrt{\frac{16\pi}{5}}n^{2}(r)\delta_{l2}
\end{eqnarray}
We note $A_{l-1}^l$,$A_{l+1}^l$,$B_{l-1}^l$ and $B_{l+1}^l$ the
coupling coefficients.

\begin{equation}
\left \{
\begin{array}{ccl}
A^{l}_{l+1}=\frac{1}{(l+1)}\frac{1}{\sqrt{(2l+1)(2l+3)}} \quad ; \quad
B^{l}_{l+1}=\frac{l(l+1)(l+2)}{\sqrt{(2l+1)(2l+3)}}\\
A^{l}_{l-1}=\frac{1}{l}\frac{1}{\sqrt{(2l-1)(2l+1)}} \quad ; \quad
B^{l}_{l-1}=\frac{l(l^2-1)}{\sqrt{(2l-1)(2l+1)}}
\end{array}
\right.
\end{equation}

Noting that the forcing implies equatorially symmetric solutions (the resulting
differential rotation is equatorially symmetric), the radial functions $w^{l}$
are non-zero only for odd $l$ while $u^{l}$ and $t^{l}$ only for even $l$. The series is
therefore $w^1, u^2, t^2, w^3, \ldots$

\end{document}